\documentclass[10pt,a4paper,reprint,english,aip,jcp,showpacs]{revtex4-1}
\usepackage[T1]{fontenc}
\usepackage[latin9]{inputenc}
\usepackage{color}
\usepackage{verbatim}
\usepackage{amsmath}
\usepackage{graphicx}
\usepackage{amssymb}
\usepackage{esint}
\usepackage{tabularx}

\makeatletter

\pdfpageheight\paperheight
\pdfpagewidth\paperwidth
\@ifundefined{definecolor}
 {\usepackage{color}}{}
\newcommand{\ket}[1]{\vert{#1}\rangle}
\newcommand{\bra}[1]{\langle{#1}\vert}
\newcommand{\rr}{{\bf r}}
\providecommand{\tabularnewline}{\\}

\makeatother

\usepackage{babel}

\begin{document}

\title{\textcolor{blue}{\Large Calculating Dispersion Interactions using
Maximally Localised Wannier Functions}}

\author{Lampros~Andrinopoulos}

\author{Nicholas~D.~M.~Hine}

\author{Arash~A.~Mostofi}

\affiliation{The~Thomas~Young~Centre for Theory~and~Simulation~of~Materials,
Imperial~College London, London SW7 2AZ, UK}

\date{\today}
\begin{abstract}
We investigate a recently developed approach\cite{silvestrelli_van_2008,silvestrelli_van_2009}
that uses maximally-localized Wannier functions (MLWFs) to evaluate
the van der Waals (vdW) contribution to the total energy of a system
calculated with density-functional theory (DFT). We test it on a set
of atomic and molecular dimers of increasing complexity (argon, methane,
ethene, benzene, phthalocyanine, and copper phthalocyanine) and demonstrate
that the method, as originally proposed, has a number of shortcomings
that hamper its predictive power. In order to overcome these problems,
we have developed and implemented a number of improvements to the
method and show that these modifications give rise to calculated binding
energies and equilibrium geometries that are in closer agreement to
results of quantum-chemical coupled-cluster calculations.\\
\mbox{ }\\
Published as \emph{J. Chem. Phys.} \textbf{135}, 154105 (2011).  
\end{abstract}


\maketitle
%
{}

\section{Introduction}

Local and semi-local exchange-correlation functionals used in density-functional
theory \cite{hohenberg_inhomogeneous_1964,kohn_self-consistent_1965}
(DFT) can not account for the effect of long-ranged dispersion, or
van der Waals (vdW), interactions. Dispersion interactions are crucial
for weakly-bound systems, particularly where no covalent or ionic
bonding is present, and often dominate intermolecular binding energies
and equilibrium geometries. Incorporating vdW interactions in DFT
remains a challenging task and a wide variety of methods have been
developed, approaching the problem from many different perspectives
\cite{zaremba_van_1976,lundqvist_density_1995,andersson_van_1996,dobson_constraint_1996,kohn_van_1998,dobson_successful_1999,rydberg_tractable_2000,dion_van_2004,tkatchenko_accurate_2009}.
In this work we focus on the method recently proposed by Silvestrelli
\cite{silvestrelli_van_2008,silvestrelli_van_2009}, which has been
recently applied to various systems \cite{Silvestrelli2009285,doi:10.1021/jp906024e,doi:10.1021/jp106627z,silvestrelli:074702}
and implemented in a number of modern electronic structure codes \cite{giannozzi_quantum_2009,Gonze-abinit-2009}.
This approach uses maximally-localized Wannier functions \cite{marzari_maximally_1997}
(MLWFs) as a means of decomposing the electronic density of the system
into a set of localized but overlapping fragments, which may then
be used to calculate a vdW correction to the DFT total energy by considering
pairwise interactions between density fragments as derived by Andersson,
Langreth and Lundqvist \cite{andersson_van_1996} (ALL).

In this Article, we explore the parameters and approximations involved
in Silvestrelli's method and improve its results where possible by
modifying various aspects of the method. We apply the method and our
proposed modifications to a series of test systems, then to two more
challenging systems, a phthalocyanine and a copper phthalocyanine
dimer. We thus demonstrate that although this method can offer an
easily implementable and computationally efficient way of calculating
the dispersion correction to the energy with the possibility of improved
accuracy (once some modifications are applied to it), it is largely
dependent on a number of parameters and choices one can make.

The remainder of the Article is organized as follows: in Sec.~\ref{sec:Methods}
we recap the necessary background theory relating to MLWFs and Silvestrelli's
method; in Sec.~\ref{sec:Improvements} we highlight some of the
problems with the method as it stands, and describe our improvements;
in Sec.~\ref{sec:Applications} we then present and discuss results
for vdW-corrected total energies and equilibrium geometries obtained
by applying these methods to a series of dimer systems and compare
to quantum chemical coupled-cluster and semi-empirical vdW (DFT+D)
approaches; finally, in Sec.~\ref{sec:Conclusions} we draw our conclusions.

\section{Theoretical Background\label{sec:Methods}}

\subsection{Maximally-Localized Wannier Functions}

Wannier functions \cite{PhysRev.52.191} are orthogonal localized
functions that span the same space as the eigenstates of a single
particle Hamiltonian. Consider the set of $N_{{\rm occ}}$ occupied
(valence) eigenstates $\{|u_{m}\rangle\}$ of a molecule. The total
energy is invariant with respect to unitary transformations among
the eigenstates \begin{equation}
\ket{w_{n}}=\sum_{m=1}^{N_{{\rm occ}}}U_{mn}\ket{u_{m}}.\label{wannier_transform}\end{equation}
 If the unitary matrix $\mathbf{U}$ is chosen such that the resulting
$N_{{\rm occ}}$ orbitals $\{w_{n}(\rr)\}$ minimize their total quadratic
spread, given by \begin{equation}
\Omega=\sum_{n}\left(\langle w_{n}\vert r^{2}\vert w_{n}\rangle-\langle w_{n}\vert\rr\vert w_{n}\rangle^{2}\right)=\sum_{n}\left(\langle r^{2}\rangle_{n}-\bar{\rr}_{n}^{2}\right),\label{SPREAD}\end{equation}
 then they are said to be maximally-localized Wannier functions~\cite{marzari_maximally_1997}
(MLWFs). Each MLWF is characterized by a value for its quadratic spread,
$S_{n}^{2}$, and its centre, $\bar{\rr}_{n}$.

In the construction of MLWFs it is sometimes useful to consider not
only the valence manifold but also a range of unoccupied eigenstates
above the Fermi level --- often those constituting the antibonding
counterparts to the valence states. This not only allows the MLWFs
to be more localized \cite{wfpart1,wfpart2} but can also restore
symmetries that would otherwise be broken arbitrarily through the
construction of MLWFs for the valence manifold only.

In order to do so, one defines an outer energy window, $E_{{\rm win}}$,
consisting of 
$N_{{\rm win}} \ge N_{{\rm occ}}$ 
states, from which one may extract an optimal $N_{{\rm dis}}$-dimensional subspace 
($N_{{\rm win}} \ge N_{{\rm dis}} \ge N_{{\rm occ}}$)
using the disentanglement approach described in Ref.~\onlinecite{souza_maximally_2001},
\begin{equation}
\ket{u_{m}^{{\rm opt}}}=\sum_{p=1}^{N_{{\rm win}}}U_{pm}^{{\rm dis}}\ket{u_{p}},\label{optimal-subspace}\end{equation}
 where $\mathbf{U}^{{\rm dis}}$ is a rectangular $N_{{\rm win}}\times N_{{\rm dis}}$
unitary matrix. $N_{{\rm dis}}$ MLWFs may then be localized by suitable
rotation of the optimal subspace in the usual manner: 
\begin{equation}
\ket{w_{n}^{{\rm dis}}}=\sum_{m=1}^{N_{{\rm dis}}}U_{mn}\ket{u_{m}^{{\rm opt}}}.\label{disentangled-mlwfs}
\end{equation}
 or, in terms of the Bloch states: 
\begin{equation}
\ket{w_{n}^{{\rm dis}}}=\sum_{m=1}^{N_{{\rm dis}}}\sum_{p=1}^{N_{{\rm win}}}U_{mn}U_{pm}^{{\rm dis}}\ket{u_{p}}.
\label{disentangled-mlwfs-bloch}
\end{equation}
 Furthermore, an inner, or \emph{frozen}, energy window may be defined
if one wishes to make certain that a range of low-lying eigenstates
are included in the optimal subspace, 
for example, the occupied states.
Algorithms for determining
MLWFs from the eigenstates obtained from electronic structure calculations
are implemented within the Wannier90 software package \cite{mostofi_wannier90:_2008}.

The single-particle density operator is given by
\begin{equation}
\hat{\rho} = \sum_{1}^{N_{{\rm occ}}} \ket{u_n}\bra{u_n}.
\label{spdo-bloch}
\end{equation} 
It can also be written in terms of
the $N_{{\rm occ}}$ fully-occupied valence MLWFs, $\ket{w_{n}}$
or equivalently in terms of a larger set of $N_{{\rm dis}}$ disentangled
MLWFs $\ket{w_{n}^{{\rm dis}}}$ 
that span the occupied subspace, which can be 
guaranteed by using a suitable frozen/inner window in the disentanglement 
procedure, and that have 
occupancies $f_{kl}^{w}$,
\begin{align}
\hat{\rho} & =\sum_{n=1}^{N_{{\rm occ}}}\ket{w_{n}}\bra{w_{n}} ,\\
& =\sum_{k,l=1}^{N_{{\rm dis}}}f_{kl}^{w}\ket{w_{k}^{{\rm dis}}}\bra{w_{l}^{{\rm dis}}},
\end{align}
where we have substituted Eq.~(\ref{wannier_transform}) 
and Eq.~(\ref{disentangled-mlwfs-bloch}), respectively, into 
Eq.~(\ref{spdo-bloch}), and
where the occupancies are given by
\begin{equation}
f_{kl}^{w}=\sum_{p=1}^{N_{{\rm occ}}}\sum_{m,s=1}^{N_{{\rm dis}}}U_{ml}U_{pm}^{{\rm dis}}U_{sk}^{*}U_{ps}^{*{\rm dis}}.
\label{wannier_occupations}
\end{equation}
 
We can write the density as a sum of diagonal ($l=k$) and off-diagonal ($l\ne k$) terms,
\begin{align}
\rho(\rr) & =\sum_{l=1}^{N_{{\rm dis}}}f_{ll}^{w}\vert w_{l}^{{\rm dis}}(\rr)\vert^{2}+\sum_{l\neq m}^{N_{{\rm dis}}}f_{lm}^{w}w_{l}^{*{\rm dis}}(\rr)w_{m}^{{\rm dis}}(\rr), 
\nonumber \\
 & \equiv \rho_{D}(\rr)+\rho_{OD}(\rr).
\label{density-dod}
\end{align}

It is important to note that in this form, 
$\rho_{D}(\rr)$ 
alone integrates to the number of valence electrons $N_{e}$,
because the mututal orthogonality of the MLWFs ensures $\int \rho_{\rm OD}(\rr) d\rr = 0$.

In the case of considering MLWFs obtained 
from the manifold of occupied states only ($N_{\rm dis}=N_{\rm occ}$), 
the the occupancy matrix is simply the identity matrix, 
$f_{kl}=\delta_{kl}$, and the charge density in terms of the MLWFs 
is simply given by
\begin{equation}
\label{density_val}
\rho(\rr) = \sum_{n=1}^{N_{\rm occ}} \vert w_n(\rr) \vert^2
\end{equation}
It is worth noting that in the case of spin-degenerate systems, the occupancies must be scaled by a factor of 2.

We have adapted the Wannier90 code to calculate the occupation matrices,
and can choose to make a \emph{diagonal} approximation to the density
by retaining only the first term of Eq.~(\ref{density-dod}). The
effect of approximating the true density with the diagonal approximation
will be discussed later in Sec.~\ref{sec:approximations_to_density} in 
the context of 
the improvements, described in Sec.~\ref{sec:Improvements}, 
to Silvestrelli's method.

\subsection{Silvestrelli's method}

Silvestrelli's approach \cite{silvestrelli_van_2008,silvestrelli_van_2009}
is based on the Andersson, Langreth and Lundqvist \cite{andersson_van_1996}
(ALL) expression for the vdW energy in terms of pairwise interactions
between density fragments $\rho_{n}(\rr)$ and $\rho_{l}(\rr')$,
separated by a distance $r_{nl}$, 
\begin{equation}
E_{{\rm vdW}}=-\sum_{n>l}g_{nl}(r_{nl})\frac{C_{6nl}}{{r_{nl}^{6}}},
\label{VDW}
\end{equation}
 where $g_{nl}(r_{nl})$ is a damping function\cite{silvestrelli_van_2009}
which screens the unphysical divergence of Eq.~(\ref{VDW}) at short
range, and \begin{equation}
C_{6nl}=\frac{3}{4(4\pi)^{3/2}}\int_{V}d\rr\int_{V'}d\rr'\frac{\sqrt{\rho_{n}(\rr)\rho_{l}(\rr')}}{\sqrt{\rho_{n}(\rr)}+\sqrt{\rho_{l}(\rr')}},\label{ALL}\end{equation}
 in atomic units. It should be noted that these expressions are only
strictly valid in the limit of non-overlapping density fragments.
There are various forms for the damping function\cite{grimme_density_2007,wu:8748}
that might have a slight short-range effect but should not affect
the long-range behaviour of the vdW energies. Here we chose to use
the damping function as proposed in the original paper by Silvestrelli\cite{silvestrelli_van_2008}.

Now, in accord with Eq.~(\ref{density_val}), 
the MLWFs obtained from the valence orbitals of a system provide
a localized decomposition of the electronic charge density, such that
$\rho_{n}(\rr)=|w_{n}(\rr)|^{2}$, so that Eq.~(\ref{ALL}) becomes
\begin{equation}
C_{6nl}=\frac{3}{32\pi^{3/2}}\int_{|\rr|\leq r_{c}}d\rr\int_{|\rr'|\leq r_{c}'}d\rr'\frac{|w_{n}(\rr)||w_{l}(\rr')|}{|w_{n}(\rr)|+|w_{l}(\rr')|},\label{eq:c6_wn_wl}\end{equation}
 where $r_{c}$ is a suitably chosen cutoff radius obtained by equating
the length scale for density change to the electron gas screening
length \cite{silvestrelli_van_2009}; we will revisit this point later.

In order to make the calculation of the integrals more tractable,
the charge density is approximated by replacing each MLWF $w_{n}(\mathbf{r})$
with a hydrogenic $s$-orbital that has the same centre $\bar{\rr}_{n}$
and spread $S_{n}$ as the MLWF, and whose analytic form is 
is given by 
\begin{equation}
w^{H}_{n}(\rr)=\frac{3^{3/4}}{\sqrt{\pi}S_{n}^{3/2}}e^{-\sqrt{3}|\mathbf{r}-\bar{\rr}_{n}|/S_{n}},\label{hyd-s}\end{equation}
 which, on substitution into Eq.~(\ref{eq:c6_wn_wl}) and after some
algebra, gives \begin{equation}
C_{6nl}=\frac{S_{n}^{3/2}S_{l}^{3}}{2\cdot3^{5/4}}F(S_{n},S_{l}),\label{c6silv}\end{equation}
 where 
\begin{equation}
F(S_{n},S_{l})=\int_{0}^{x_{c}}dx\int_{0}^{y_{c}}dy\frac{x^{2}y^{2}e^{-x}e^{-y}}{e^{-x}/\beta+e^{-y}},
\label{f_integral}
\end{equation}
 $\beta=(S_{n}/S_{l})^{3/2}$, $x_{c}=\sqrt{3}r_{c}/S_{n}$ and $y_{c}=\sqrt{3}r_{c}'/S_{l}$.
Whereas evaluating Eq.~(\ref{eq:c6_wn_wl}) using the 
true MLWFs requires a computationally demanding six-dimensional 
numerical integration,
Eq.~(\ref{f_integral}) may be evaluated easily since it 
is only a two-dimensional integral that
depends solely on the MLWF spreads and centres, not their detailed shapes
or orientations. 

We note that in the case of 
a spin-degenerate system, since
every MLWF is doubly occupied, the density of each fragment must be multiplied
by a factor of 2 and, therefore, the $C_{6nl}$ integral in Eq.~(\ref{eq:c6_wn_wl}) must be scaled by a factor of $\sqrt{2}$.

\section{Improvements to Silvestrelli's Method}

\label{sec:Improvements}

The approximations that go into the method described in the previous
Section will clearly not always hold, and the need to examine them
is clear. In this Section, we introduce our enhancements to the method
that address possible drawbacks.

\subsection{Partly Occupied Wannier Functions}

Using a manifold of eigenstates that includes but is larger than the
subspace spanned by just the valence states results in partly-occupied
MLWFs that are generally more localized and that better reflect the
symmetries of the system, as opposed to MLWFs obtained by rotation
of the valence subspace only, which arbitrarily break the symmetry
(we will demonstrate examples of this phenomenon in Sec.~\ref{sec:Applications}).

In order to account for the partial occupancy of the MLWFs, we make
a slight modification to Silvestrelli's approach, explicitly introducing
occupancies in the definition of the $C_{6nl}$ integral; since in
the diagonal approximation, the density of each fragment is now given
by $\rho_{n}(\rr)=f_{nn}^{w}|w_{n}(\rr)|^{2}$, the expression for
$F(S_{n},S_{l})$ in Eq.~(\ref{f_integral}) becomes 
\begin{equation}
F(S_{n},S_{l})=\int_{0}^{x_{c}}dx\int_{0}^{y_{c}}dy\frac{x^{2}y^{2}e^{-x}e^{-y}}{e^{-x}/(\beta\sqrt{f_{nn}^{w}})+e^{-y}/\sqrt{f_{ll}^{w}}},\label{f_integral_occupations}
\end{equation}
 where the $f_{nn}^{w}$ are given by Eq.~(\ref{wannier_occupations}).
We will see in Sec.~\ref{sec:Applications} that this seemingly simple
idea can give rise to a marked improvement in the accuracy of the
method.

\subsection{Modification to describe $p$-like states}

\begin{figure}
\includegraphics[clip,width=0.3\columnwidth]{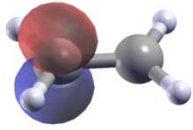} \caption{\label{fig:ethene_pz}Partly occupied $p$-like orbital on ethene
molecule. In the method described here, each of the two lobes (coloured
red and blue) is replaced by an $s$ orbital and considered a separate
fragment.}

\end{figure}

MLWFs describing only the valence manifold often take the form of
well-localized functions centred on a bond between two atoms, and
are thus reasonably well-described by the approximation of replacing
them with a suitable $s$-orbital. When anti-bonding states are included
in the construction of the MLWFs, the resulting orbitals have more
atomic-orbital character. This is demonstrated by the atom-centred
$p$-like MLWF shown in Fig.~\ref{fig:ethene_pz}.
It is clear that the density associated with such an MLWF will not
be very well represented by a single $s$-like function at its centre.
In order to approximate $p$-like orbitals appropriately when calculating
$C_{6}$, one could imagine using a suitably-oriented analytic expression
for a hydrogenic $p$-orbital, for example, a canonical $p_{z}$-orbital
given by \begin{equation}
p_{z}(\rr)=\frac{30^{5/4}r\cos{\theta}}{\sqrt{32\pi}S^{5/2}}e^{-\sqrt{30}r/2S},\label{hyd-pz}\end{equation}
 which has been normalized such that its quadratic spread is $\langle p_{z}|(\rr-\bar{\rr})^{2}|p_{z}\rangle=S^{2}$.
As a consequence of the explicit angular dependence, using this function
in Eq.~(\ref{eq:c6_wn_wl}) would give rise to four-dimensional integrals
for which analytic solutions are not readily available. Numerical
evaluation of these integrals, for realistic systems, would be prohibitively
computationally expensive. We solve this problem by identifying the
$p$-like MLWFs in the system and replacing them with the hydrogenic
form given in Eq.~(\ref{hyd-pz}). Then, we further approximate each
lobe (lower and upper) of this $p$-like orbital with two separate
hydrogenic $s$-orbitals of the form of Eq.~(\ref{hyd-s}). In order
to do so, 
for each of the upper ($+$) and lower ($-$) lobes 
of the orbital,
it is necessary to know 
the spread $S_{\pm}$ and centre
$\bar{\rr}_{\pm}$,
given by 
\begin{align}
S_{\pm}^{2} & =\int_{0}^{\infty}\int_{0}^{\pi/2}\int_{0}^{2\pi}r^{4}p_{z}^{2}(\rr)\sin\theta drd\theta d\phi,\\
\bar{\rr}_{\pm} & =\bar{\rr}\hspace{1mm}\pm\int_{0}^{\infty}\int_{0}^{\pi/2}\int_{0}^{2\pi}r^{3}\cos\theta\, p_{z}^{2}(\rr)\sin\theta drd\theta d\phi\:\hat{{\bf z}},\end{align}
 which, after some algebra, simplifies to 
\begin{align}
S_{\pm} & =\frac{7S}{8\sqrt{2}},\label{spreads_pz}\\
\bar{\rr}_{\pm} & =\bar{\rr}{}\pm\frac{15S}{8\sqrt{30}}\:\hat{{\bf z}},\label{centres_pz}
\end{align}
 where $\bar{\rr}$ and $S$ are the original centre and spread, respectively,
of the true MLWF. 
These expressions may be easily generalized to 
arbitrary orientations of the symmetry axis of a $p$-like state 
by rotating the offset vectors $(\bar{\rr}_{\pm}-\bar{\rr})$ accordingly.

Thus, we have developed a formalism whereby the charge density due
to MLWFs with $p$-like character can be represented by a pair of
$s$-like hydrogenic orbitals with appropriate centres and spreads.
In Sec.~\ref{sec:Applications} we will show how this works in practice
for calculating vdW energy corrections.

In the relatively simple systems studied in this paper, the $p$-like
orbitals are easily distinguished from other orbitals by their partial
occupancies, given by Eq.~(\ref{wannier_occupations}), which are typically
closer to 0.5 rather than 1. Alternatively, and especially for structurally
more complex systems, the shape of each MLWF could be characterized
using the efficient method described in Appendix A of Ref.~\onlinecite{shelley}
as another means of automating the procedure of identifying $p$-like
functions.

\subsection{Symmetry Considerations}

Minimizing the total spread $\Omega$ with respect to the elements
of the unitary matrix $\mathbf{U}$, and thus producing MLWFs, has
the effect of picking from the space of all possible unitary matrices
one which produces the most localized Wannier functions accessible
through optimization from a chosen initial guess. This is often enough
to uniquely determine the MLWFs. In some cases, however, it does not
give rise to a unique choice, even if the optimization procedure is
perfect. For example, the atomic positions and electron density of
the system may possess certain symmetry elements, such as rotations
about a particular axis. Then there will exist a number of equally
valid and degenerate representations of the MLWFs and their centres,
which give the same spread, and are related by symmetry. The minimization
procedure breaks the symmetry by choosing one of these representations;
in other words there will be a degree of arbitrariness in the final
MLWFs. It is clear from Eq.~(\ref{VDW}) that any degree of non-uniqueness
of the centres will cause an undesirable variability of the vdW energy
calculated in Silvestrelli's method. This is indeed what we observe
in some of the examples below. Moving away from a description of the
MLWFs using the valence states only, and towards using partly occupied
MLWFs that include anti-bonding states and which retain the symmetries
of the system, enables us to overcome these problems, as we demonstrate
below.

\section{Applications\label{sec:Applications}}

\subsection{Calculation Details}

For the application of Silvestrelli's method to the following dimer
systems we used the Quantum Espresso (QE) package\cite{giannozzi_quantum_2009}
to perform the ground-state DFT calculations, and Wannier90\cite{mostofi_wannier90:_2008}
to obtain the centres and spreads of the MLWFs.
Our results are compared to both the semi-empirical DFT+D method\cite{grimme_semiempirical_2006,JCC:JCC21112}
as implemented in QE, which is expected to give good asymptotic behaviour,
and a wavefunction-based coupled-cluster approach, CCSD(T), which
is considered the `gold-standard' of quantum chemistry.

The PBE~\cite{perdew_generalized_1996} generalized-gradient approximation
for exchange and correlation, except in the case of argon where the
revPBE~\cite{PhysRevLett.80.890} functional was used; norm-conserving
pseudopotentials, and $\Gamma$-point sampling of the Brillouin zone
were used throughout. We note that we have chosen to use revPBE for
the argon system since PBE produces significant binding in rare gas
dimers as it overestimates the long-range part of the exchange contribution\cite{dion_van_2004,PhysRevLett.95.109902,PhysRevLett.96.146107}.
For all the other systems we studied in this manuscript, however,
PBE does not cause spurious binding and would therefore normally be
considered an appropriate functional. A plane-wave basis set cut-off
energy of 80~Ry was used in all calculations with QE except for the
case of the phthalocyanine and copper phthalocyanine where a 50~Ry
energy cutoff was used. %
For the dimers of argon, methane, ethene, phthalocyanine and copper
phthalocyanine, cubic simulation cells of length 15.87~\AA{}, 15.87~\AA{},
21.16~\AA{}\ and 23.81~\AA{}, respectively, were used. For the
dimers of benzene, a hexagonal cell with $a=15.87$~\AA{}\ and $c=31.75$~\AA{}\ was
used. For all the systems, the choice of energy windows when using
the disentanglement procedure in Wannier90 for our modified method
was as follows: inner (frozen) energy windows were chosen to include
all the valence states; outer energy windows ranged from the lowest
eigenvalue of the system, $\epsilon_{0}$, to a maximum of $E_{{\rm win}}=\epsilon_{{\rm LUMO}}+\alpha(\epsilon_{{\rm HOMO}}-\epsilon_{0})$,
where $\epsilon_{{\rm HOMO}}$ is the energy of the highest occupied
valence Kohn-Sham (KS) state and $\epsilon_{{\rm LUMO}}$ is the energy
of the lowest unoccupied KS state. The factor~$\alpha=0.4$ was chosen
to scale down the valence energy bandwidth, used to estimate the energy
difference required above the LUMO when including anti-bonding states.
We discuss the sensitivity of the method to this factor in Sec.~\ref{sec:sensitivity_to_window}.

\subsection{Argon}

We will first investigate the severity of the aforementioned issues
relating to symmetry, by considering the case of an argon dimer. Optimization
of the MLWFs describing a single argon atom produces four doubly occupied
MLWFs arranged tetrahedrally around the atom. Due to spherical symmetry,
the orientation of these MLWFs with respect to a given coordinate
system is arbitrary for an isolated atom and the final MLWFs obtained
will depend on the initial guess used. In the dimer, this arbitrariness
is removed, at least in principle, since the spherical symmetry is
broken by the presence of the other atom at a specific orientation.
At large separations, this is not in practice necessarily the case:
the electron density overlap between the Ar atoms is vanishingly small,
since the wavefunctions decay exponentially away from the atom. Therefore,
to within attainable numerical precision, the orientation of the MLWFs
on each atom is uncorrelated with the orientation of the other atom:
the MLWFs can be freely rotated with respect to the atom without affecting
the total spread. Note, however, that since the vdW energy only decays
as $R^{-6}$, its value \textit{is} influenced by the orientation
of the MLWF centres (and hence their separation) out to distances
beyond which the calculated spread (and thus the optimised MLWF orientation)
has ceased to be sensitive to separation.

\begin{figure}
\includegraphics[clip,scale=0.3]{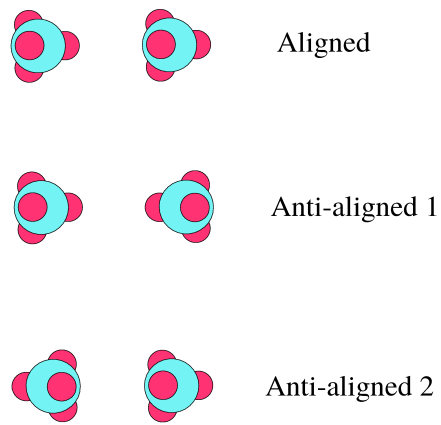} \caption{\label{fig:argon_orientations} Illustration of three of the many
possible configurations of MLWF centres (small pink spheres) for the
two argon atoms (large blue spheres) in the fragment method.}

\end{figure}

\begin{figure}
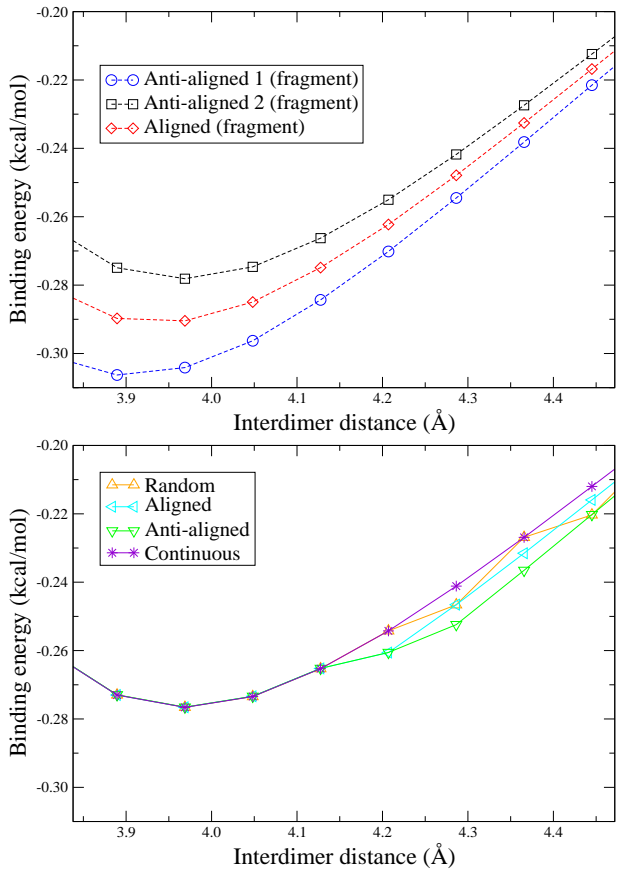

\includegraphics[clip,scale=0.33]{ARGON_FRAG}

\includegraphics[clip,scale=0.33]{ARGON_FULL} \caption{\label{fig:argon_binding} Binding energy versus interatomic separation
for the argon dimer, for varying relative orientations of the MLWF
centres surrounding each atom (see Fig.~\ref{fig:argon_orientations}).
Top panel: results obtained using the fragment method, in which the
MLWF centres are calculated for a lone Ar atom and then translated
and rotated to the second Ar atom. Bottom panel: results obtained
using the true MLWF centres with various initial guesses for their
positions. The curve labelled `continuous' is obtained by using the
MLWF centres from a configuration at small separation as the initial
guess for the centres at larger separations. In this way, the discontinuities
in the curve are avoided and a unique curve is obtained (see text
for details). }

\end{figure}

This dependence can be investigated in a two-atom system by fixing
the relative orientations of the MLWF centres between the two atoms
in the dimer. This is achieved by first calculating the MLWF centres
for a single atom of argon and then translating and rotating these
centres to the second Ar atom with various choices of alignment. We
will refer to this approach as the \emph{fragment method}. In this
method, we calculate the dispersion correction to the energy for a
dimer system using various possible arrangements of MLWF centres on
the other atom. Three possible high-symmetry choices are shown in
Fig.~\ref{fig:argon_orientations}. For each of these orientations,
Fig.~\ref{fig:argon_binding} (top) shows the binding energy of the
Ar dimer as the separation of the atoms varies. We see that there
is considerable displacement of the curves, and the binding energy
and the equilibrium separation change according to the alignment chosen
by up to $0.04$~kcal/mol and 0.08~\AA{}, respectively.

In contrast to this fragment approach, in Fig.~\ref{fig:argon_binding}
(bottom) we show the binding energy as calculated with the normal
approach of using the optimized MLWFs of the entire dimer system.
However, here we have used varying initial guesses corresponding to
the set of possible alignments shown in Fig.~\ref{fig:argon_orientations}.
We see that at small separations, the MLWF centres always converge
to the same positions, regardless of the initial guess, and the binding
energy curve is nearly independent of the choice of initial guess
($\sim10^{-3}$~kcal/mol variation).

At larger separation, however, the spread minimization is insufficiently
sensitive to the relative orientation of the MLWFs on different atoms,
and does not necessarily alter it from the initial guess, resulting
in several different possible results depending on the initial orientation
of the centres. If a random initial guess is chosen, then the energy
varies discontinuously, as a function of separation, within the bounds
imposed by the limiting cases described using the fragment method.
This is because the MLWF centres converge to different orientations
depending on their starting positions (curve labelled `random' in
Fig.~\ref{fig:argon_binding} (bottom)).

In order to avoid this problem of non-uniqueness of binding energy
curves, a random initial guess is used first for a configuration at
small separation, in the knowledge that the result will be independent
of the guess used. Then the centres computed at the previous, smaller
separation are used as the initial guess for the calculation at a
larger separation. In this manner, a unique continuous curve is obtained
(labelled `continuous' in Fig.~\ref{fig:argon_binding} (bottom)).
This is the approach that we adopt for all subsequent calculations
in this paper.

From the continuous curve, we obtain 3.97~\AA{}\ for the equilibrium
separation and $-0.28$~kcal/mol for the binding energy. This is
in good agreement with the coupled cluster CCSD(T) calculations of
Ref.~\onlinecite{slavicek:2102}, which give 3.78~\AA{}\ and
$-0.28$~kcal/mol, respectively, whereas revPBE without dispersion
corrections gives 4.62~\AA{}\ and $-$0.04~kcal/mol.

\subsection{Methane}

\begin{figure}
\includegraphics[clip,scale=0.23]{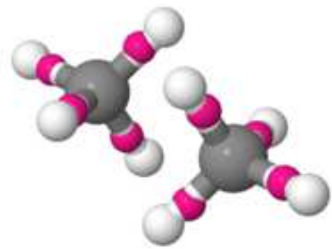} \caption{\label{fig:methane_pic} Illustration of the methane dimer. Carbon
atoms are shown by large grey spheres, hydrogen by small white spheres,
and the valence MLWF centres are shown by small pink spheres.}

\end{figure}

The methane dimer is a straightforward application of the Silvestrelli
method: the positions of the MLWF centres, which lie on the four tetrahedral
C-H bonds of each CH$_{4}$ molecule (see Fig.~\ref{fig:methane_pic}),
obey the same symmetries as the atomic positions, so there exists
no arbitrariness of orientation.

\begin{figure}
\includegraphics[clip,scale=0.33]{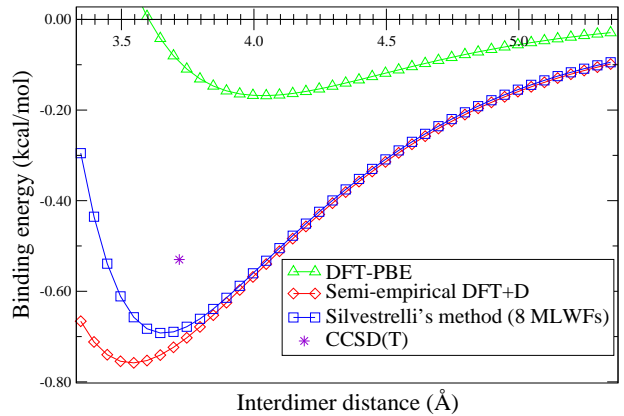} \caption{\label{methane_binding}Binding energy curves for the methane dimer
with various methods.}

\end{figure}

In Fig.~\ref{methane_binding}, we compare to the results of both
DFT+D and CCSD(T) calculations. Our geometries and CCSD(T) results
were drawn from the Benchmark Energy and Geometry Database (BEGDB)\cite{hobza_begdb}.

The accuracy of Silvestrelli's method in the case of the methane dimer
is good compared to CCSD(T): the former gives an equilibrium separation
of 3.66~\AA{}\ and binding energy of $-0.69$~kcal/mol, and the
latter 3.72~\AA{}\ and $-0.53$~kcal/mol, respectively. DFT+D is
in somewhat worse agreement with CCSD(T), yielding 3.54~\AA{}\ and
$-0.76$~kcal/mol, respectively.

\subsection{Ethene}

We now turn our attention to the ethene dimer, which includes a C-C
double bond. Again we will compare results for the original and
modified methods against CCSD(T) and DFT+D results. We have again used
the geometries for each molecule taken from the BEGDB.

\begin{figure}
\includegraphics[clip,scale=0.28]{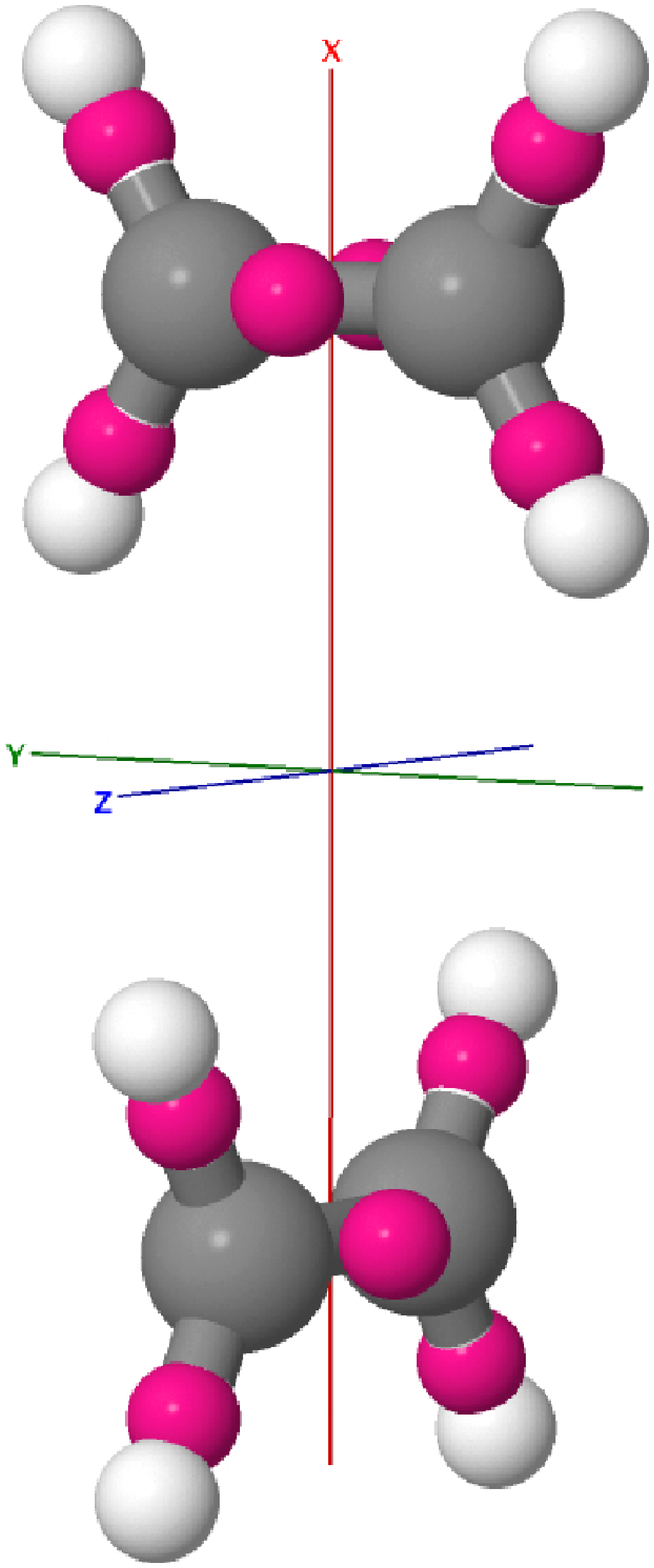} \includegraphics[clip,scale=0.28]{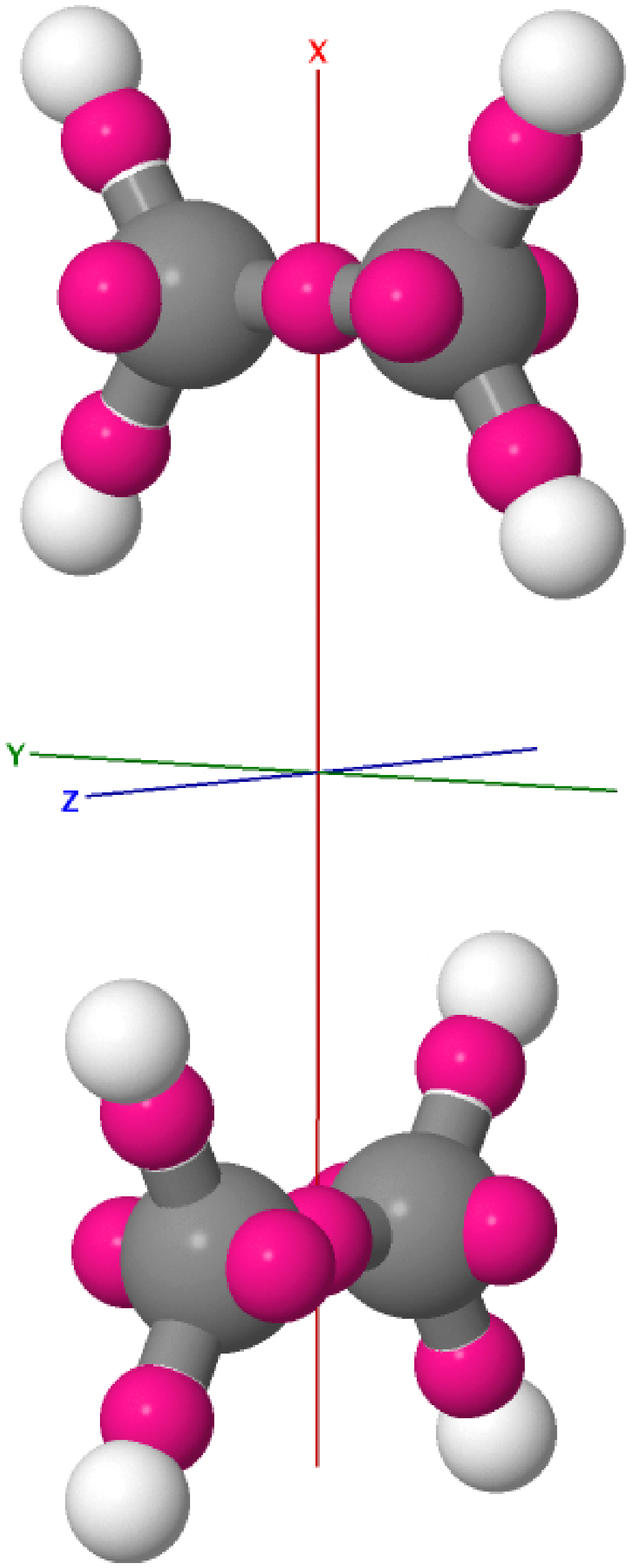}
\caption{\label{fig:ethene_dimer}Colours as in Fig.~\ref{fig:methane_pic}.
Left: Ethene dimer with six MLWFs per molecule. Right: Ethene dimer
with seven MLWFs per molecule. The centres of the $p$-like MLWFs
are placed on the carbon atoms, but here we show the centres of the
individual lobes of these $p$-like orbitals as calculated by our
method.}

\end{figure}

To use Silvestrelli's original method in this case, we include only
the valence manifold in the creation of the MLWFs, giving
six MLWFs per molecule arranged as shown in Fig.~\ref{fig:ethene_dimer} (left).
In our modified method we use seven MLWFs per molecule, with $p$-like,
partly occupied orbitals on each carbon atom (Fig.~\ref{fig:ethene_dimer}
(right)). %

\begin{figure}
\label{ETHENE} \includegraphics[clip,scale=0.33]{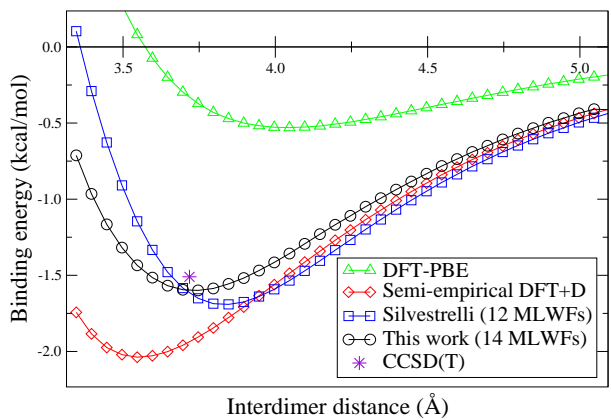}
\caption{\label{fig:ethene_binding}Binding energy for an ethene dimer with
various methods.}

\end{figure}

As seen in Fig.~\ref{fig:ethene_binding}, neither the original Silvestrelli
method (blue squares) nor DFT+D (red diamonds) reproduce the CCSD(T)
values very accurately. By expanding the manifold of eigenstates used
in the construction of the MLWFs and applying our modified method to
include partial MLWF occupancies and splitting of the $p$-like functions
(see Sec.~\ref{sec:Improvements}), we find an excellent agreement
(black circles) with the CCSD(T) equilibrium values of 3.72~\AA{}\ for
the separation and $-1.51$~kcal/mol for the binding energy; our
method gives 3.73~\AA{}\ and $-1.60$~kcal/mol, respectively; Silvestrelli's
method gives 3.83~\AA{}\
and $-1.69$~kcal/mol; DFT+D yields 3.55~\AA{}\ and $-2.04$~kcal/mol.

\subsection{Benzene}

\begin{figure}
\includegraphics[width=0.95\columnwidth]{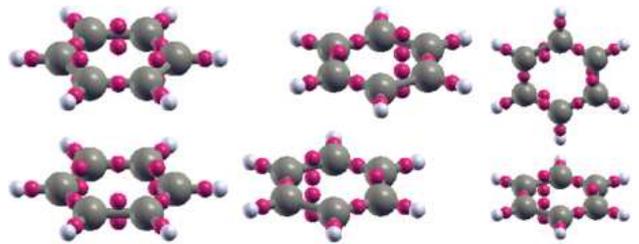} \caption{\label{fig:benzene_pic}The three configurations used for the benzene
dimer calculations: S (vertical displacement), PD (vertical and lateral
displacement) and T (vertical displacement plus rotation in plane
of one molecule), and the valence MLWF centres in each case (depicted
by pink spheres).}

\end{figure}

For benzene, the valence states can be represented by 15 doubly-occupied
Wannier functions. The MLWF optimization procedure in this case therefore 
breaks the $D_{6h}$ symmetry of the benzene ring: the end result is that
there are three C-C `double' bonds and three C-C `single' bonds in the MLWF
representation. Those alternating double and single C-C bonds represent 
a delocalised $\pi$-bond around the ring. The double bonds are represented
by two centres located above and below the plane of the molecule,
while the single bonds are represented by one centre on the bond.
When two molecules are put in proximity (see Fig.~\ref{fig:benzene_pic})
and the vdW energy is calculated by Silvestrelli's method, the breaking
of the symmetry affects the vdW energy in an arbitrary manner, dependent
on how the two rings are aligned (i.e. whether the pairs of double
bonds in adjacent molecules are aligned or anti-aligned). This alignment
is defined by where the initial guesses for the centres of the Wannier
functions are placed.

The case of the benzene dimer therefore illustrates again the need
to include the unoccupied antibonding states in the construction of
the MLWFs: doing so increases the number of MLWFs to 18 and introduces
partial occupancies, but restores the $D_{6h}$ symmetry of the system
and also localises the MLWFs more. This then makes the vdW contribution
independent of the initial guess for the Wannier function centres.

\begin{figure}
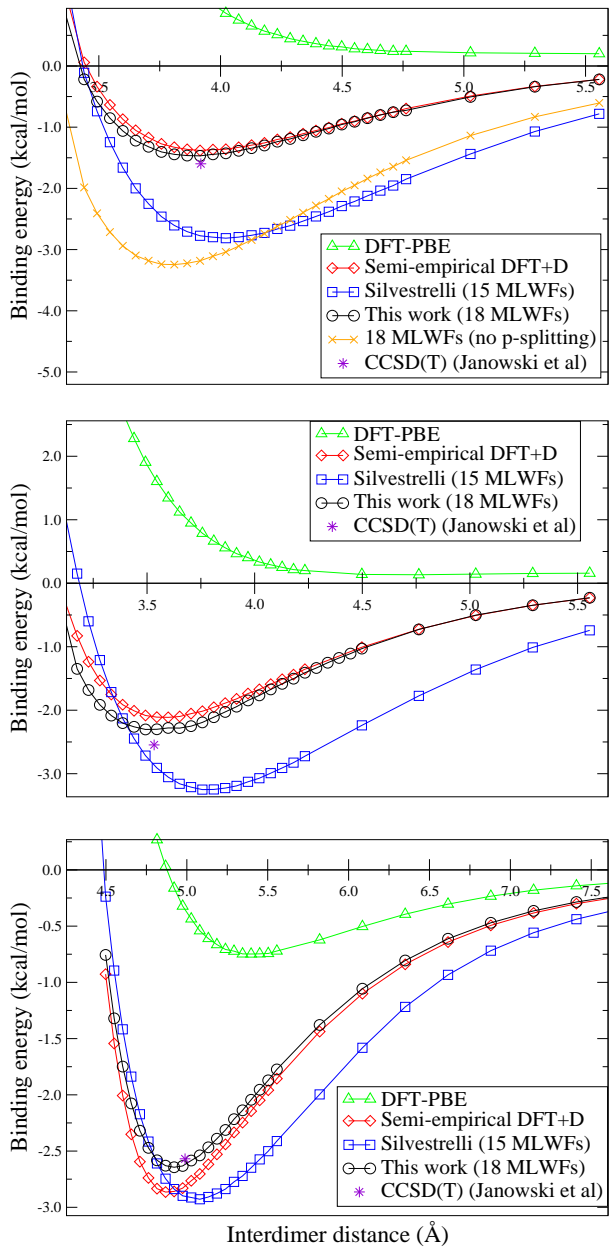

\includegraphics[clip,scale=0.33]{S}

\includegraphics[clip,scale=0.33]{PD}

\includegraphics[clip,scale=0.33]{T} \caption{\label{benzene_binding}Binding energy (kcal/mol) curves for the various
methods for the benzene dimer in the S, PD and T configurations (top,
middle and bottom respectively). For the S configuration we also show
the curve using 18 MLWFs per molecule if no $p$-splitting is used;
in this case the method overbinds. CCSD(T) benchmark values are from
Janowski \emph{et al.}~\cite{Janowski200727} }

\end{figure}

We applied our implementation of the original Silvestrelli's method
(with 15 MLWFs), and then our modified method (with 18 MLWFs, partial
occupancies and splitting of $p$-like states) to determine the binding
energy as a function of displacement for three types of displacement
(labelled S, PD, and T, illustrated in Fig.~\ref{fig:benzene_pic}
of one of the molecules in the benzene dimer. We compare this to DFT+D
and to the CCSD(T) calculations of Ref.~\onlinecite{Janowski200727}.
We note that we used the same bond lengths for C-C and C-H as Ref.~\onlinecite{Janowski200727}
to within two decimal places, to construct perfectly symmetric benzene
rings for our calculations.

The binding energy curves for the various methods for the three configurations
are shown in Fig.~\ref{benzene_binding}. Silvestrelli's method (blue squares)
does not agree very well with CCSD(T) calculations, overestimating equilibrium
distances by 0.07-0.25~\AA{}\ (Table~\ref{table-distance})
and overestimating binding energies by 0.28-1.25~kcal/mol (Table~\ref{table-energy}).
In particular, the dispersion curve obtained from Silvestrelli's method
does not agree asymptotically with the DFT+D curve (red diamonds).
In the T configuration Silvestrelli's method performs better in terms
of equilibrium distance, binding energy and asymptotics as it can
be seen in Fig.~\ref{benzene_binding} (bottom).

For the S configuration we also show the binding curve obtained if
the anti-bonding states are included in the construction of the MLWFs,
but splitting of the $p$-like states is not used (orange crosses);
it is clear that in this case the method does not perform well, as
replacing a $p$-like orbital by an $s$-orbital is a very poor approximation.

Our full modified method, including both the larger manifold and the
splitting of $p$-like states (black circles in Fig.~\ref{benzene_binding}),
on the other hand, has excellent agreement in terms of equilibrium
distances and binding energies with the DFT+D curves and the CCSD(T)
values, for all three configurations, to within 0.05~\AA{}\ and
0.33~kcal/mol (Table~\ref{table-distance} and~\ref{table-energy});
the asymptotic behaviour of the energy is also better captured.

\begin{table}
\begin{tabular}{|c|c|c|c|}
\hline 
Method  & S  & PD  & T \tabularnewline
\hline 
Silvestrelli (15 MLWFs)  & 4.01  & 3.78  & 5.06 \tabularnewline
This work (18 MLWFs)  & 3.89  & 3.55  & 4.88 \tabularnewline
Semi-empirical DFT+D  & 3.93  & 3.58  & 4.89 \tabularnewline
CCSD(T) (Janowski \textit{et al} \cite{Janowski200727})  & 3.92  & 3.53  & 4.99\tabularnewline
\hline
\end{tabular}\caption{\label{table-distance}Equilibrium distances in \AA{}\ for the benzene
dimers in the three configurations (Fig.~\ref{fig:benzene_pic})
using the various methods. For all DFT calculations the PBE functional
was used.}

\end{table}

\begin{table}
\begin{tabular}{|c|c|c|c|}
\hline 
Method  & S  & PD  & T \tabularnewline
\hline 
Silvestrelli (15 MLWFs)  & $-$2.85  & $-$3.23  & $-$2.85 \tabularnewline
This work (18 MLWFs)  & $-$1.47  & $-$2.31  & $-$2.64 \tabularnewline
Semi-empirical DFT+D  & $-$1.38  & $-$2.11  & $-$2.87 \tabularnewline
CCSD(T) (Janowski \textit{et al} \cite{Janowski200727})  & $-1.60$  & $-2.55$  & $-2.57$\tabularnewline
\hline
\end{tabular}\caption{\label{table-energy}Binding energies (kcal/mol) at equilibrium geometry
for the benzene dimers in the three configurations (Fig.~\ref{fig:benzene_pic})
using the various methods. For all DFT calculations the PBE functional
was used.}

\end{table}

\subsection{H$_{2}$Pc and CuPc}

\begin{figure}
\includegraphics[clip,scale=0.17]{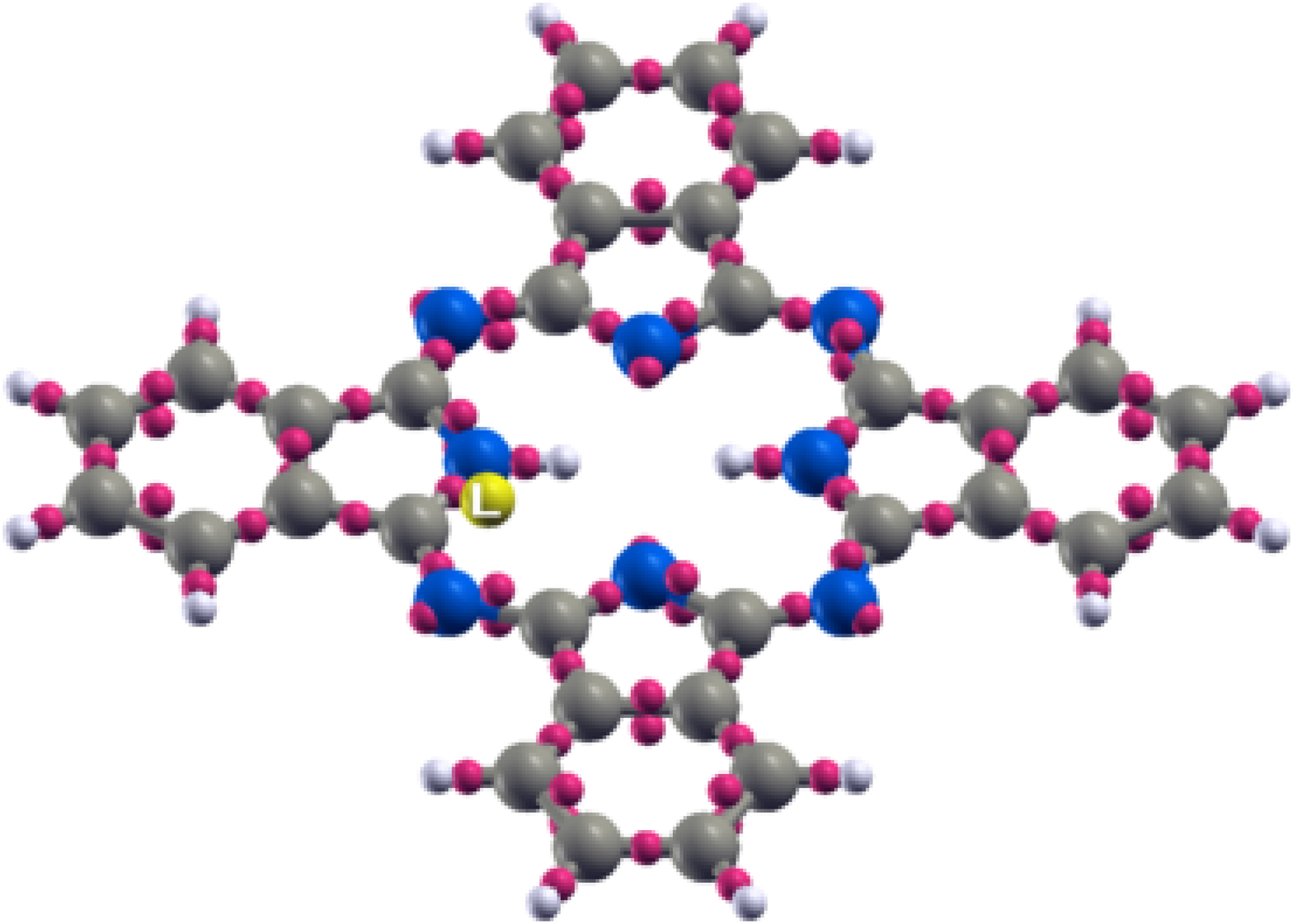} 
\vskip1cm \includegraphics[clip,scale=0.17]{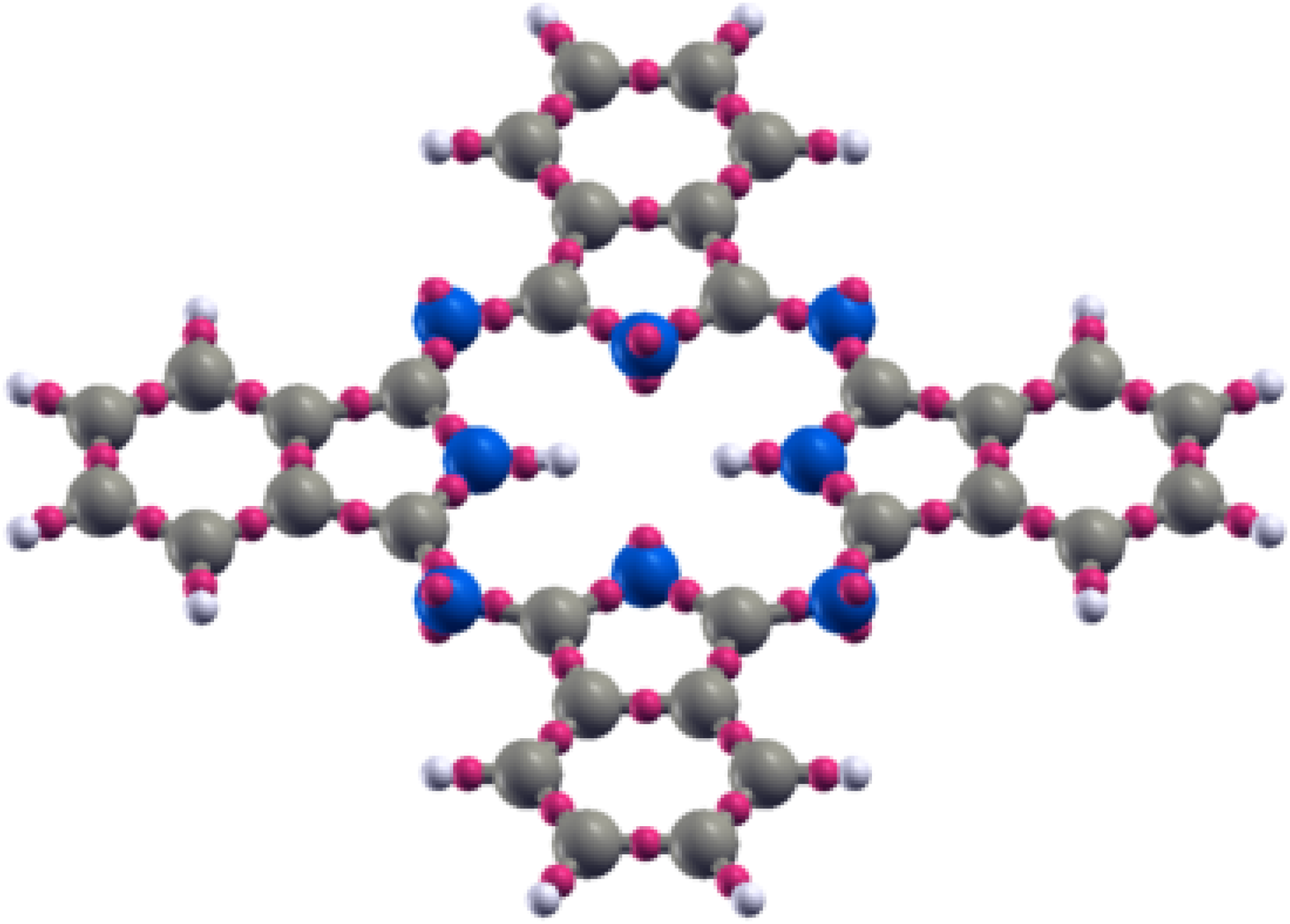}
\caption{\label{Pcs}Left: Phthalocyanine ($\text{H}_{2}\text{Pc}$) molecule
and its valence MLWF centres. Hydrogen atoms are by small white spheres,
carbon atoms by large grey spheres and nitrogen atoms by large blue
spheres. The MLWF centres are shown by the small pink spheres. Using
only the valence MLWFs does not give a satisfactory description of
the system since it yields a lone MLWF of unphysically large spread
(shown by large yellow sphere and labelled by the letter L). Right:
H$_{2}$Pc molecule and its 112 MLWF centres, now including anti-bonding
states. With this representation all the $D_{2h}$ symmetry of the
ring is restored and a better chemical picture is given. There are
$s$-like orbitals on every bond and the non-hydrogenated nitrogens,
and $p$-like partly occupied orbitals on every carbon the two hydrogenated
nitrogens (not shown here as these are located inside the corresponding
atoms).}

\end{figure}

To examine the difficulties encountered applying these methods to
larger systems, we have investigated the phthalocyanine (H$_{2}$Pc)
dimer in the simplest configuration (S vertically displaced) first
by applying Silvestrelli's method and then by applying our modifications
it, and comparing the binding energy curve to one obtained using DFT+D.
The optimised MLWF centres for a single H$_{2}$Pc are shown in Fig.~\ref{Pcs}
(top). We see that as with the benzene molecule, there are alternating
single and double MWLF centres on the C-C bonds of the six-membered
rings, representing delocalised $\pi$-bonds. We also find, however,
that using only the 93 valence MLWFs (186 valence electrons) is problematic,
as a good representation of the electronic density of the system cannot
be obtained in this way since this breaks the symmetry of the system,
but most importantly it yields one lone MLWF of unrealistically large
spread ($\sim$2.5~\AA{}) located some distance from any atoms (Fig.~\ref{Pcs}
(top)). This is due to the fact that an odd number (93 MLWFs) is incompatible
with the $D_{2h}$ symmetry of the molecule.

\begin{figure}
\includegraphics[clip,scale=0.33]{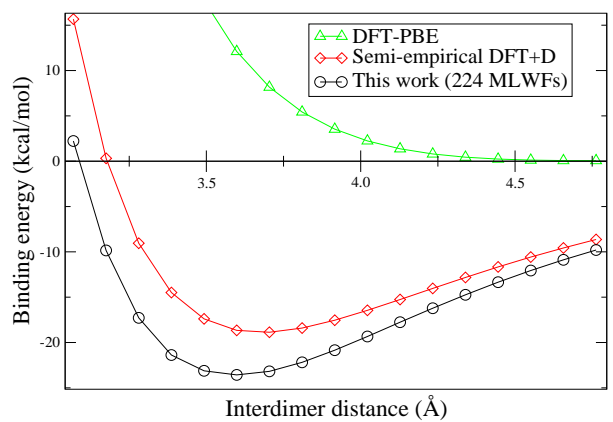} \caption{\label{h2pc_binding}Binding energy curves for H$_{2}$Pc dimer in
the S configuration (vertically displaced) versus intermolecular distance
obtained with the various methods.}

\end{figure}

Using a larger and even number of MLWFs (112 per molecule) we can
restore this $D_{2h}$ symmetry of the molecule (Fig.~\ref{Pcs}
(bottom)) and represent the electronic density of the system in a
way more compatible with its chemistry. When anti-bonding states are
included, it is important to make a chemically intuitive initial guess
for the centres and forms of the MLWFs. We make initial guesses as
follows: we place $p$-like orbitals on the carbon atoms and $s$-like
orbitals on every bond and $p$-like orbitals on the hydrogenated
nitrogens as well as two $s$-like orbitals on every non-hydrogenated
nitrogen atom. In this way, we have partly occupied MLWFs that represent
the 372 valence electrons of the dimer. The binding energy curves
obtained by using this representation and our modifications to Silvestrelli's
method are shown in Fig.~\ref{h2pc_binding} and compared to DFT+D.
The binding energy obtained from our method is $-23.63$~kcal/mol
and the equilibrium distance 3.58~\AA{}; with DFT+D we obtain $-18.91$~kcal/mol
and 3.68~\AA{}. As for benzene, we see very good agreement with DFT+D;
these values roughly agree with the stacking distance of crystalline
H$_{2}$Pc (around 3.2--3.4~\AA{})\cite{orti:1009}. Silvestrelli's
original method severely overbinds the dimer (giving a binding energy
of $-41$~kcal/mol) because of the unphysically large spread of the
lone MLWF that appears in the valence representation. This is due
to the strong dependence of the vdW energy on the spreads (Eq.~(\ref{c6silv})).

\begin{figure}
\includegraphics[clip,scale=0.17]{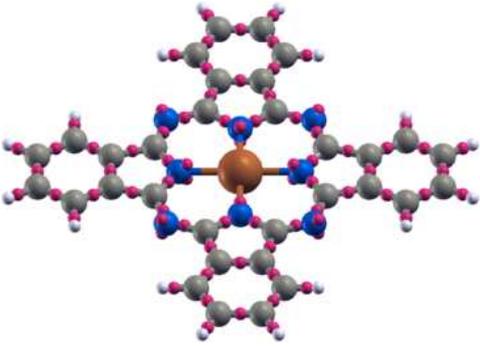}
\caption{\label{CUPC}Copper phthalocyanine (CuPc) molecule and its 234 MLWF
centres, again including anti-bonding states. Colours as in Fig.~\ref{Pcs},
with copper shown by the large brown sphere in the centre. There are
$s$-symmetry MLWFs on every bond and atom except for copper, $p$-like
MLWFs on the carbons and 5 $d$-symmetry MLWFs on the copper atom.
Now there are no $p$-like orbitals on any nitrogen atom as for H$_{2}$Pc.}

\end{figure}

In the case of CuPc dimer (vertically displaced S configuration) we
again do not use the valence manifold of 390~MLWFs per dimer (195~MLWFs
per molecule: 98 spin up and 97 spin down), but instead use a larger
manifold of MLWFs. We note that the dimer configuration used here
does not correspond to any phases CuPc is observed in experimentally,
but was used for illustrative purposes as it is the simplest one.
This is a spin-polarized system, so a different set of MLWFs is required
for spin up/down electrons, yielding a total of 234 singly occupied
MLWFs per molecule (117 for every spin channel). There are 10 $d$-like
MLWFs (five for every spin channel) centred on each copper atom, and
$s$-like MLWFs on bonds and nitrogens. The MLWFs corresponding to
spin up and spin down electrons have essentially the same centres
for the same bonds or atoms (Fig.~\ref{CUPC}).

In such cases, where some Wannier functions centres are very closely
centred, it would be incorrect to consider them as separate fragments
since this would violate the fundamental assumption of the ALL method,
that it is valid for non-overlapping fragments only. This can be understood
from the fact that Eq.~(\ref{ALL}) is strongly non-linear, so adding
the contributions of overlapping density fragments does not give the
same result as summing the densities beforehand. As a result, Silvestrelli's
method severely overbinds the dimer ($\sim-108$~kcal/mol), demonstrating
that the method breaks down for overlapping fragments.

We alleviate this problem by amalgamating all the centres and spreads
of the closely placed MLWFs (in this case the $d$-like MLWFs on Cu)
into one MLWF with a centre and spread given by the arithmetic mean
of the closely placed MLWFs, and occupancies given by the sum of the
separate MLWFs. The criterion for amalgamating MLWFs can be automated
such that MLWFs less than a particular threshold distance apart are
combined. In our case, we used a value of 0.1~\AA{}\ for this threshold,
which had the desired effect of including the $d$-like orbitals on
Cu in the amalgamation procedure, while leaving all other MLWFs in
the system unaffected.

\begin{figure}
\includegraphics[clip,scale=0.33]{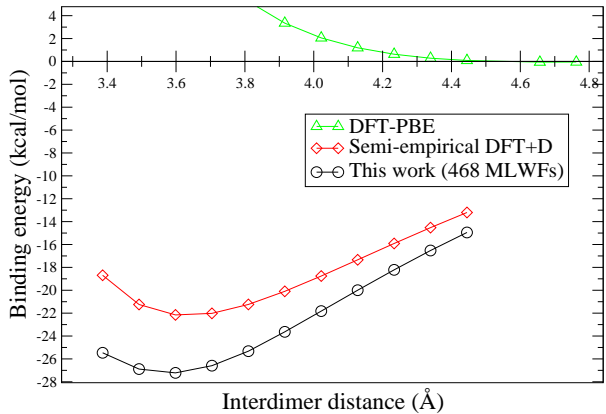}\caption{\label{cupc_binding}Binding energy curves
for the CuPc dimer in the S configuration (vertically displaced) obtained
using the various methods.}

\end{figure}

In Fig.~\ref{cupc_binding} we compare the binding energy curves
obtained using DFT+D to our modified method (now including the amalgamation
of closely-overlapping MLWFs) using a larger manifold of 468 MLWFs per
dimer. This gives much more sensible results, with a binding energy
of $-27.22$~kcal/mol and an equilibrium separation of 3.57~\AA{},
in fair agreement with DFT+D, which gives $-22.21$~kcal/mol and
3.63~\AA{}, respectively. These values are in reasonable agreement
with those for H$_{2}$Pc (as obtained using our method above),
and also with those obtained with other methods for other metal
phthalocyanines (NiPc and MgPc calculated with the TS-vdW scheme in
Ref.~\onlinecite{marom_describing_2010} using the PBE funtional).

\subsection{Intermolecular $C_{6}$ coefficients}

It is expedient to define effective intermolecular $C_{6}$ coefficients,
\begin{equation}
C_{6{\rm eff}}=\frac{1}{2}\sum_{n,l}C_{6nl},\label{c6eff}\end{equation}
 where only intermolecular terms are summed over, i.e., $n$ and $l$
correspond to MLWFs on different molecules, and the factor of $1/2$
accounts for double-counting. In Table~\ref{tab:c6eff}, we compare
our values to those of the original method of Silvestrelli, benchmark
dispersion-corrected MP2 calculations (MP2+$\Delta$vdW) and reference
results obtained using the Dipole Oscillator Strength Distribution
(DOSD) approach, given in the database of 
Ref.~\onlinecite{tkatchenko_dispersion-corrected_2009}.

As previously discussed in Ref.~\onlinecite{silvestrelli_van_2009},
comparison with reference values is made somewhat difficult by the
fact that they are obtained by fitting to experimental data and hence
also include higher-order terms ($C_{8}$, $C_{10}$) in an effective
manner.

Taking the reference values as a benchmark, it can be seen from Table~\ref{tab:c6eff}
that, for the systems under consideration, there is no clear or systematic
improvement in calculated effective $C_{6}$ coefficients with our
modifications to Silvestrelli's approach as compared to Silvestrelli's
original approach: in the case of ethene the original method compares
more favourably, while in the case of the benzene dimers our approach
performs much better. In spite of this, however, it is worth noting
that our approach (as shown earlier) significantly improves the values
obtained for equilibrium separations and binding energies, as compared
to CCSD(T), for all systems considered for which we have access to
CCSD(T) results.

\begin{table}
\begin{tabular}{|c|c|c|c|c|}
\hline 
\textbf{System}  & \multicolumn{4}{c|}{$C_{6}$ ($E_{h}a_{0}^{6}$)}\tabularnewline
\hline 
 & Silvestrelli  & This work  & MP2+$\Delta$vdW  & pseudo-DOSD \tabularnewline
\hline 
Argon  & 92.4  & 92.4  & 76.1  & 64.3 \tabularnewline
Methane  & 99.1  & 99.1  & 119  & 130 \tabularnewline
Ethene  & 275  & 261  & 328  & 300 \tabularnewline
Benzene S  & 2727  & 1288  & 2364  & 1723 \tabularnewline
Benzene PD  & 2727  & 1284  & 2364  & 1723 \tabularnewline
Benzene T  & 2769  & 1262  & 2364  & 1723 \tabularnewline
\hline
\end{tabular}\caption{Effective intermolecular $C_{6}$ coefficients. Dispersion-corrected
MP2 (MP2+$\Delta$vdW) and reference values are drawn from Ref.~\onlinecite{tkatchenko_dispersion-corrected_2009}.
For the argon and methane dimers, our approach is identical to the
original method of Silvestrelli. The differences between the values
reported in the first column (Silvestrelli) and those in Ref.~\onlinecite{silvestrelli_van_2009}
are attributable to the different calculational details such as choice
of exchange and correlation functional, simulation cell size and plane-wave
energy cutoff.}

\label{tab:c6eff} 
\end{table}

\subsection{Sensitivity to cutoff radius $r_{c}$}

\begin{figure}
\includegraphics[clip,scale=0.33]{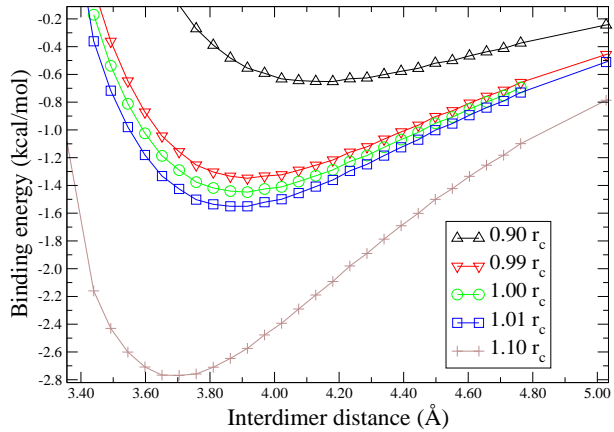}

\caption{\label{cutoff_rc}Binding energy curve for the benzene dimer in the
S configuration for various values of $r_{c}$ using our modified
method with 18 MLWFs per molecule.}

\end{figure}

The sensitivity of the binding energy on the cutoff radius $r_{c}$
in Eq.~\ref{c6silv} was tested on the S configuration of the benzene
dimer with 18 MLWFs per molecule (Fig.~\ref{cutoff_rc}). Even small
changes of 1\% in the cutoff radius result in significant changes
in the binding energy curves, with the binding energy and equilibrium
distance varying by 6-8\% and 0.2-0.8\% respectively. For larger changes
in $r_{c}$, the method breaks down, as the energy changes are unphysically
large. Although the cutoff radius is physically justified \cite{andersson_van_1996},
this strong dependence of the vdW correction on it is a weakness of
the method.

\subsection{Approximations to the density}
\label{sec:approximations_to_density}
In the original method of Silvestrelli, the KS density is approximated by replacing all real MLWFs with hydrogenic $s$ wavefunctions $w_n^H(\rr)$ given by Eq.~(\ref{hyd-s}); 
for the purpose of calulating the $C_6$ coefficients, 
the electronic charge density of the system is, therefore, effectively 
approximated as
\begin{equation}\label{density_silv}
\rho_{\rm s}(\rr) = \sum_{n=1}^{N_{\rm occ}} \vert w_n^H(\rr) \vert^2 .
\end{equation}

In the modified method presented here, in which the MLWFs are constructed using
a manifold of the KS states beyond just the occupied orbitals, there
are two levels of approximation
to the charge density.
First, the off-diagonal
component $\rho_{OD}(\rr)$ is neglected from Eq.~(\ref{density-dod}) and, 
second, the ``hydrogenic'' approximation of the original approach is 
applied, whereby the disentangled Wannier functions, $w_n^{\rm dis}(\rr)$, 
are replaced by hydrogenic orbitals, $w_n^H(\rr)$, of the same center 
and spread.
In our method, therefore, the density is approximated as
\begin{equation}\label{density_method}
\rho_{\rm dis}(\rr) = \sum_{n=1}^{N} f^w_{nn} \vert w_n^H(\rr) \vert^2,
\end{equation}
where $N$ is now the of total number of fragments, after the splitting of 
$p$-like orbitals or amalgamation of co-centric MLWFs has been performed.
We consider each of these approximations in turn for 
a typical system, the benzene molecule. 

The XCrySDen\cite{xcrysden} package was used to generate the 
isosurface plots referred to in this Section.

\begin{figure}
\includegraphics[scale=0.17]{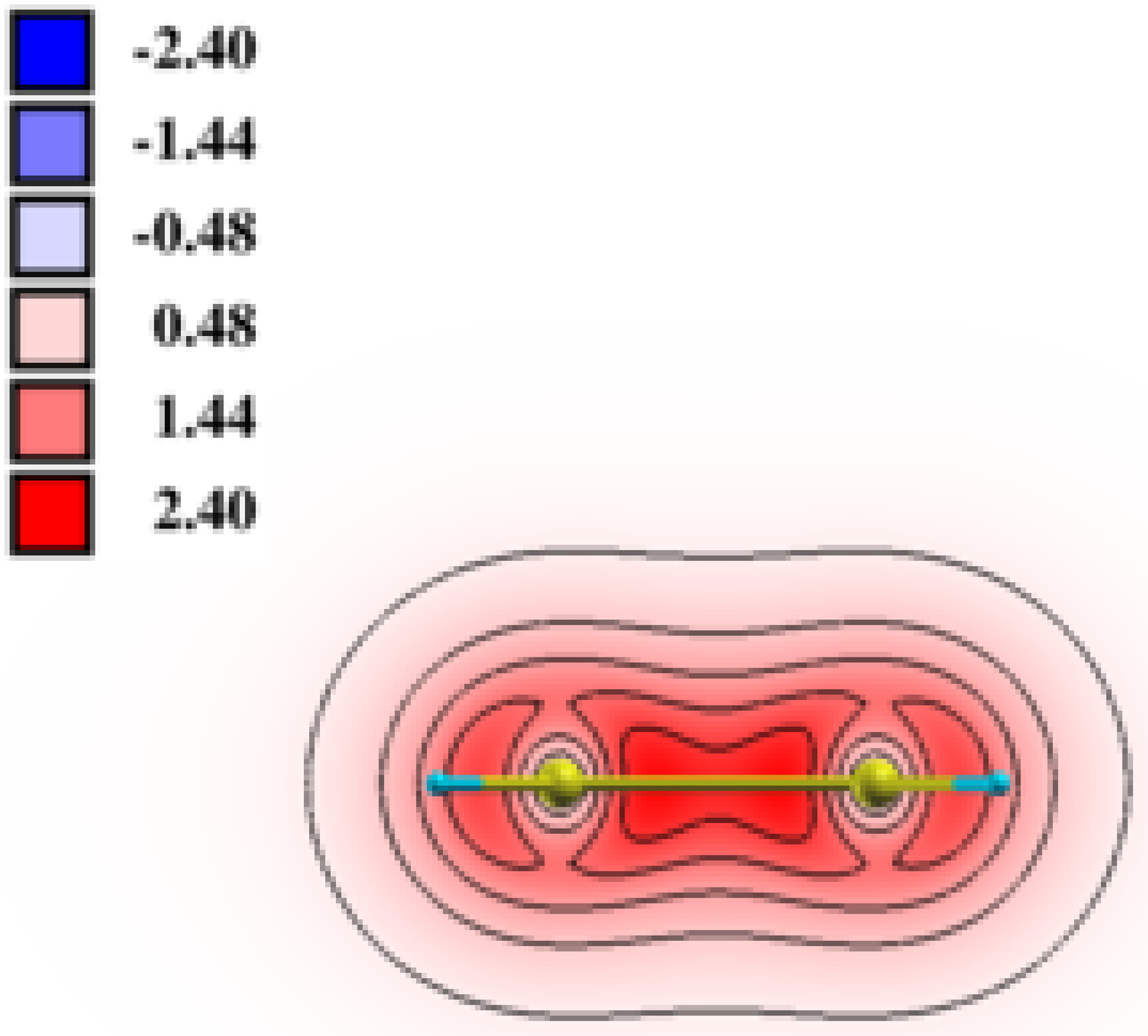} 
\includegraphics[scale=0.17]{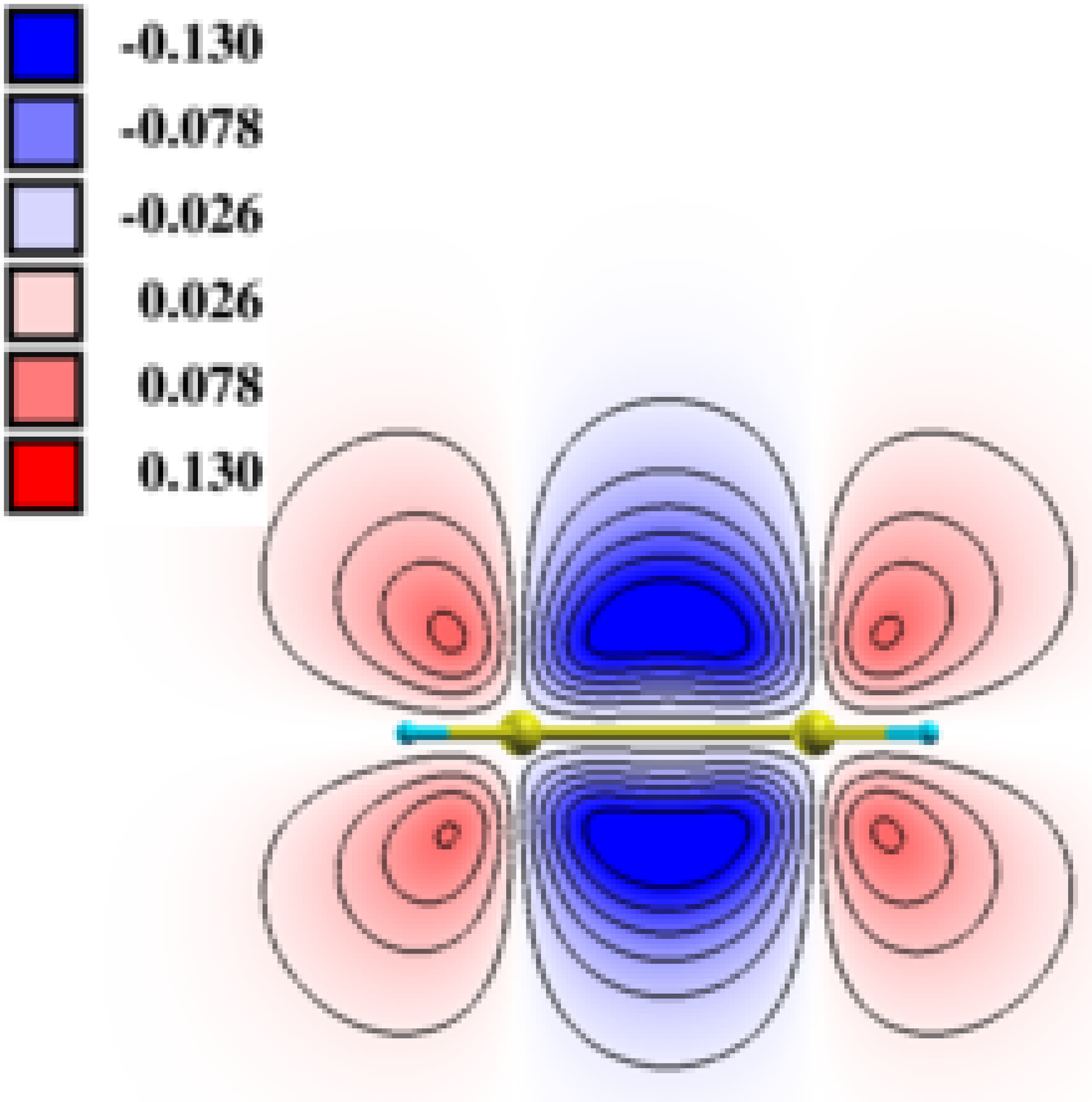}
\caption{\label{density_profile_real_od} Density profile on a plane 
parallel to a C-C bond ($xz$-plane) in a benzene molecule. 
Left: the original Kohn-Sham density $\rho(\rr)$ from the 
plane-wave DFT calculation. Right: The off-diagonal component $\rho_{OD}(\rr)$ 
of the density (see Eq.~(\ref{density-dod})) when a disentangled 
manifold is used to construct $N_{\textrm{dis}}=18$ MLWFs. Note the 
much-reduced scale compared to that of the total density. 
The units are \AA$^{-3}$.}
\end{figure}

In Fig.~\ref{density_profile_real_od} we show density isosurface 
plots for the KS density $\rho(\rr)$ (left) and the off-diagonal 
density $\rho_{\rm OD}(\rr)$ (right), which emphasises that the latter 
is uniformly small in magnitude,
comprising only a small fraction of the total density ($\sim 5-7\%$), 
as a result of the exponential localisation of the MLWFs.

\begin{figure}
\includegraphics[scale=0.17]{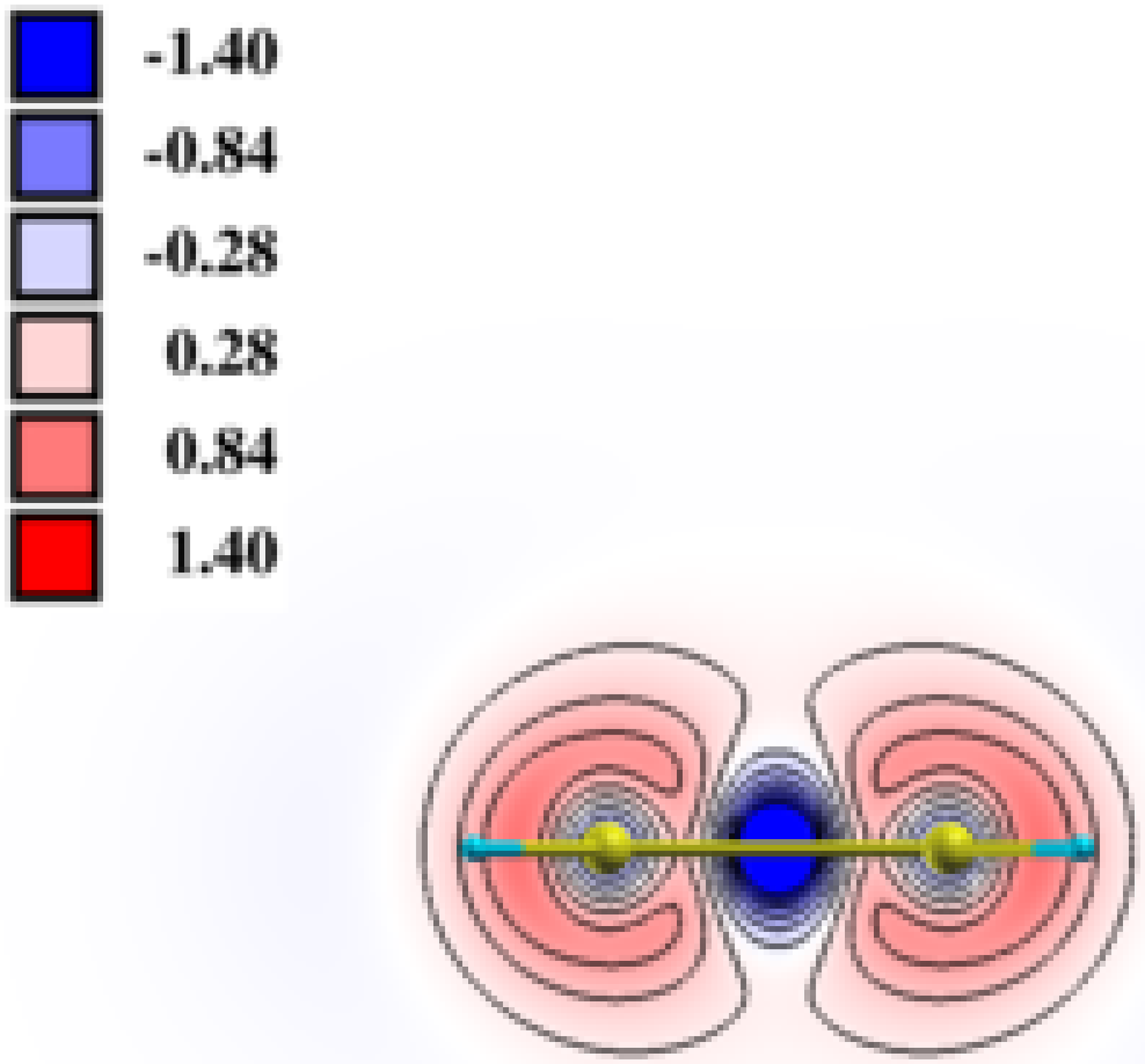} 
\includegraphics[scale=0.17]{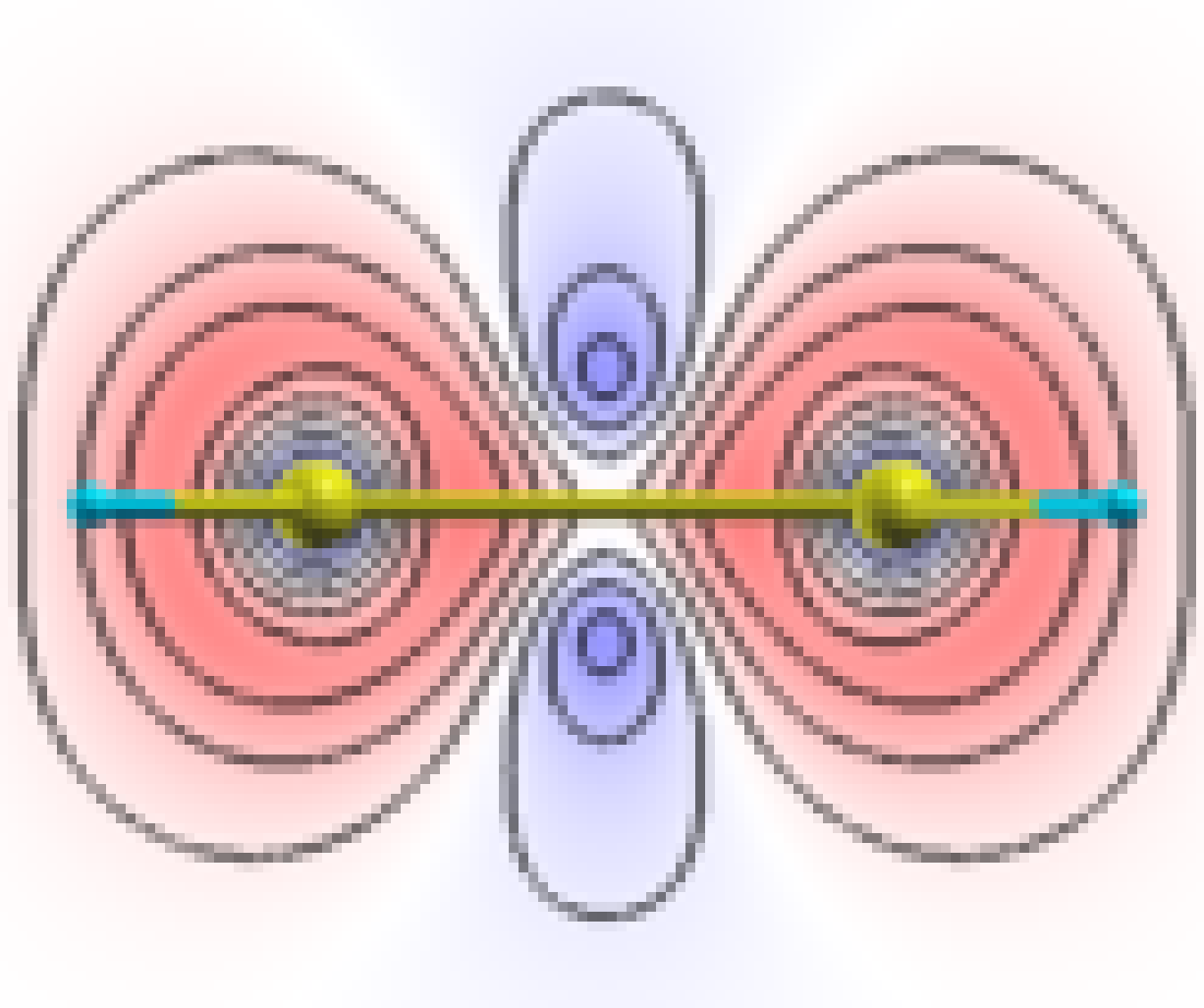} 
\caption{\label{density_profile_silv} 
Density difference isosurface plots showing the difference 
$\rho(\rr)-\rho_{\rm s}(\rr)$ between the KS density and the 
approximate ``hydrogenic'' density of the original Silvestrelli approach 
(Eq.~(\ref{density_silv})). Left: cross-section through a ``single'' bond. Right: cross-section through a ``double'' bond.
}
\end{figure}

In Fig.~\ref{density_profile_silv} we show the difference between the 
KS density $\rho(\rr)$ and the ``hydrogenic'' approximation 
$\rho_{\rm s}(\rr)$ of the original Silvestrelli method 
(Eq.~(\ref{density_silv})) for two of the C-C bonds in benzene: 
on the left a ``single'' bond; on the right a ``double'' bond. 
These two bonds only differ because of the symmetry-breaking inherent 
in the MLWF construction when just the valence states are used. 
We see that the density associated with the $\pi$-bond is not well 
represented in either case.

\begin{figure}
\includegraphics[scale=0.17]{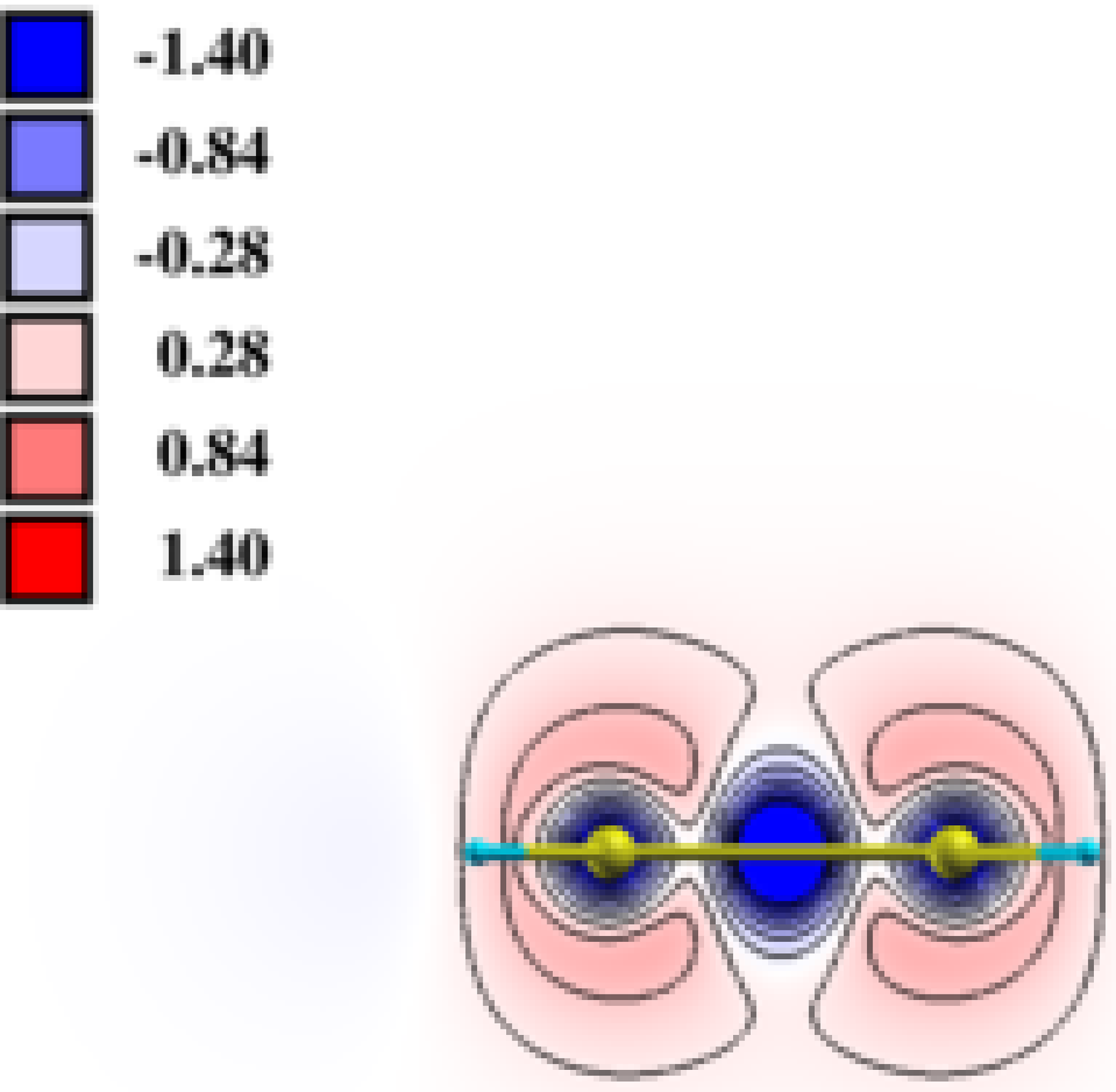}
\includegraphics[scale=0.17]{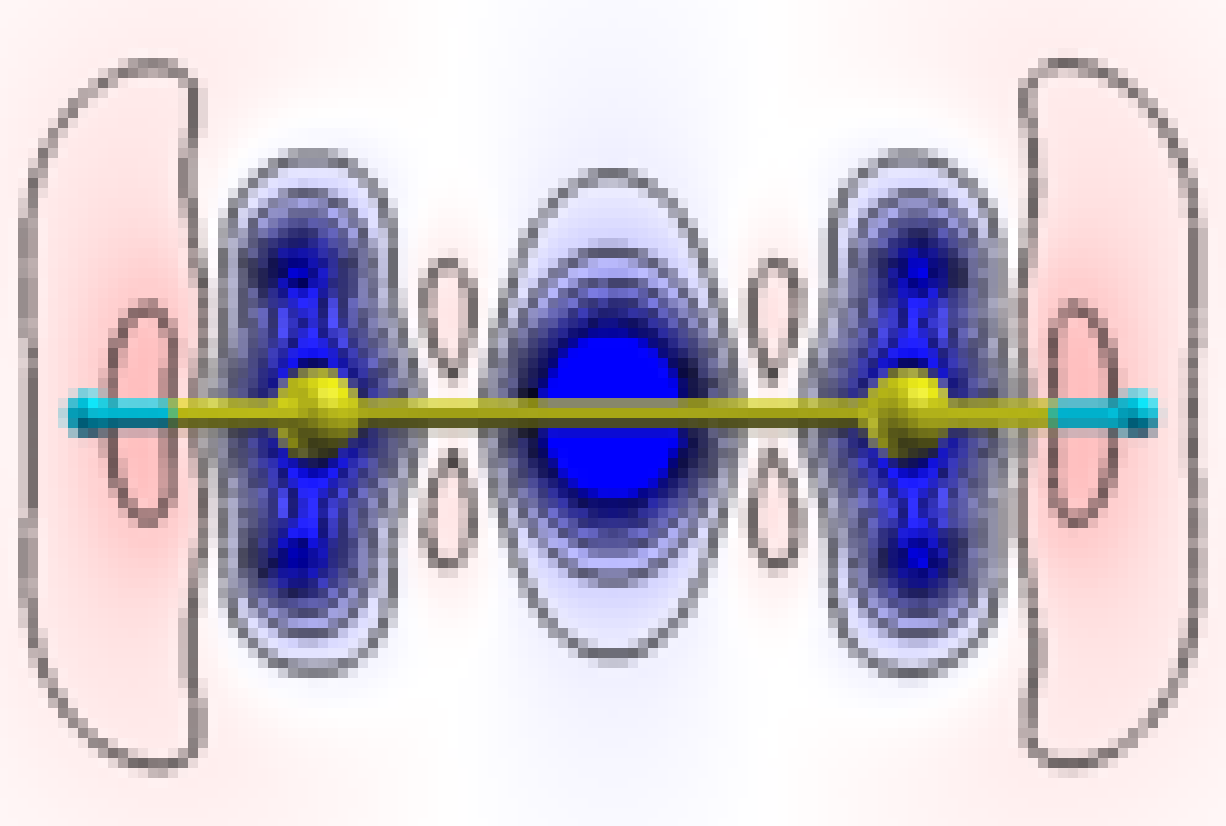}
\caption{\label{density_profile_dis} 
Density difference isosurface plots, on the same plane 
as in Fig.~\ref{density_profile_real_od}, showing the 
difference $\rho(\rr)-\rho_{\rm dis}(\rr)$ between the 
KS density and the ``hydrogenic'' density of our method 
when a disentangled set of MLWFs is used (Eq.~(\ref{density_method})). 
Left: without splitting of $p$-like states; right: with splitting 
of $p$-like states into two $s$-like states. The mean difference with the KS density 
compared to the original Silvestrelli's method is 
reduced overall for both cases, but even more in the case of $p$-splitting.}
\end{figure}

Finally, in Fig.~\ref{density_profile_dis}, we show the difference between 
the KS density $\rho(\rr)$ and that of our modified method 
$\rho_{\rm dis}(\rr)$ (Eq.~(\ref{density_method})) with 18 MLWFs 
obtained by disentanglement from a larger manifold. The left-hand 
plot is without splitting the $p$-like states, and the right-hand 
plot is with (as described in Sec.~\ref{sec:Improvements}).

We see that while this introduces small regions where the density
differs significantly (right at the MLWF centers), everywhere else
it is overall an improvement, producing a better representation of
the density compared to the original Silvestrelli's method, 
especially in the case of $p$-splitting.

In summary, discarding the off-diagonal component of the density (in the 
case of disentangled MLWFs) is a relatively minor approximation, and has 
a considerably smaller effect than approximating the 
density in various ways using hydrogenic orbitals, the latter 
being inherent to both our approach and the original approach of Silvestrelli.
The maximum difference between the KS density $\rho(\rr)$ and the density in
our method is reduced by $\sim 23\%$ and the minimum difference by $\sim 5\%$,
compared to the difference between the KS density and the density in 
Silvestrelli's method.

\subsection{Sensitivity to energy window $E_{{\rm win}}$}

\label{sec:sensitivity_to_window} To use our modifications to Silvestrelli's
method, the disentanglement procedure has to be used in the construction
of Wannier functions, as outlined in the Methods section. Because including
ever more high-energy plane-wave states inevitably allows extra variational
freedom in the construction of the MLWFs, we find that the precise values of
the MLWF spreads are sensitive to the outer energy window used for the 
disentanglement. Specifically as $E_{{\rm win}}$
is increased, the MLWFs become more localised (their spreads decrease).
As a result, the vdW energy is also affected by the choice of $E_{{\rm win}}$.

\begin{figure}
\includegraphics[clip,scale=0.33]{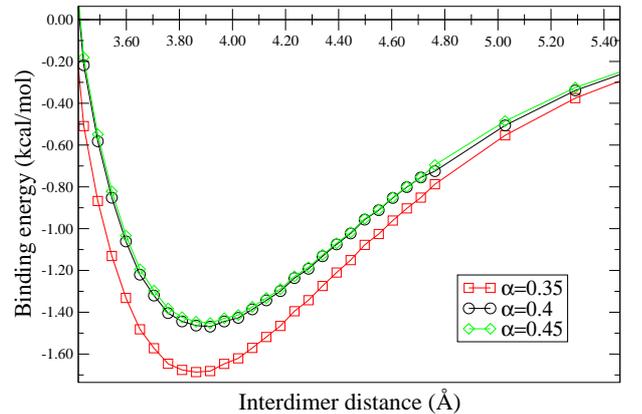}

\caption{\label{cutoff_alpha}Binding energy curve for the benzene dimer in the
S configuration for various values of $\alpha$ using our modified method
with 18 MLWFs per molecule.}

\end{figure}

In this work we have chosen throughout to estimate an appropriate energy 
window using
\begin{equation}
\label{eq:alpha_window}
E_{{\rm win}}=\epsilon_{{\rm LUMO}}+\alpha(\epsilon_{{\rm HOMO}}-\epsilon_{0})
\end{equation}
where $\alpha$ is a factor used to scale the valence energy bandwidth.
This is motivated by the idea that to enable us to restore the symmetry,
we need to include the antibonding counterparts to the valence states, without
including too large a number of irrelevant higher-lying unbound states. Eq.~(\ref{eq:alpha_window}) is an attempt to estimate the range of energies
spanned by these antibonding states.
In Fig.~\ref{cutoff_alpha} we show the dependence of the vdW binding energy 
curves for the benzene dimer in the S configuration on $\alpha$. While there
is considerable variation for too-small $\alpha$, we find that for 
values beyond 0.4, the curves vary only a rather small amount with $\alpha$; 
As long as a value of $\alpha$ around this value is chosen, it should yield reasonable
results, suggesting the extra degree of empiricism introduced by this
procedure is relatively limited in scale. The value of $\alpha$ was set 
to 0.4 in all the other calculations in this work.

\section{Conclusion}

\label{sec:Conclusions}

We conclude that Silvestrelli's method is computationally efficient
and very easy to implement for small systems where initial guesses
for the Wannier centres can be specified. However, there is a very
strong dependence of the calculated vdW energy on the position and 
spread of the Wannier centres, and these are not always as unique 
as one might hope. Symmetry-breaking, often induced by considering
only the valence manifold in the construction of the MLWFs, may 
introduce arbitary dependence on initial guesses in a way that 
significantly affect the vdW energy. We have shown that
arbitrarily-broken symmetries may often be restored by increasing
the number of Wannier functions used and generating them with a suitably-chosen
range of the conduction states as well as the valence states. This
necessitates the inclusion of occupancies in the formalism. We note
that in cases where no symmetries are restored when we use more MLWFs,
as in the example of ethene, it is the better localisation of the
MLWFs that may be responsible for improved vdW energies, since the
method is based on pairwise summation of well-separated fragments.

Particularly, in cases with a larger number of Wannier functions,
we have shown that the approximation implicit in replacing the true
Wannier functions with hydrogenic $s$-orbitals may not always yield
an accurate representation of the electronic density, and have shown
how in cases where there is $p$-like symmetry, it is better to substitute
the $p$-symmetry functions with two $s$-like functions. By considering
the problems associated with applying these adapted methods to larger
systems such as H$_{2}$Pc and CuPc, we have demonstrated that the
approach is not necessarily a good candidate for studying larger systems,
where specifying initial guesses for a large number of non-trivial
MLWFs may be difficult; chemical insight for the form of these higher-lying
states has to be employed, but becomes more difficult for even larger
systems. In the case of copper phthalocyanine, we showed that MLWFs
that are centred effectively at the same point (such as the five $d$-like
MLWFs on each Cu atom) cannot be treated as separate fragments of
density; they should instead be amalgamated into one fragment of density
of an averaged centre and spread and summed occupancies. The reason
for this is that the method is valid only in the limit of well-separated
fragments. Finally, we have demonstrated that there is also a strong
dependence of the vdW energy on the cutoff radius used in the integral
of Eq.~(\ref{c6silv}), and although the value used is justified
on physical grounds, it nevertheless represents something of an adjustable
parameter with considerable influence on the results obtained. Overall,
we conclude that while Silvestrelli's method suffers from several drawbacks,
it can be made rather accurate once modifications are applied to it (albeit
with the introduction of further empirical character); 
these improvements, and Silvestrelli's method in general, however, may be
less suitable for more structurally complex, large-scale systems, for which
alternative methods that are more fully \textit{ab initio} may be desirable.

\begin{acknowledgments}
The authors acknowledge the support of the Engineering and Physical
Sciences Research Council (EPSRC Grant No. EP/G055882/1) for funding
through the HPC Software Development program. The authors are grateful
for the computing resources provided by Imperial College's High Performance
Computing service, which has enabled all the simulations presented
here.
\end{acknowledgments}

%

\end{document}